\documentclass[12pt,oneside]{article}

\usepackage{epsfig,lscape}
\usepackage{amssymb,amsfonts,amsmath}
\usepackage{rotating}

\usepackage{amsthm}

\usepackage{color}
\usepackage[dvipsnames]{xcolor}

\usepackage{float}
\usepackage{hyperref}  

\def\me{\mathrm e}

\def\var{\mathrm{var}}

\def\N{\mathrm{N}}

\def\T{ {\mathrm{\scriptscriptstyle T}} }

\def\expit{\mathrm{expit}}
\def\plim{\mathrm{plim}}

\newtheorem{pro}{Proposition}
\newtheorem{cor}{Corollary}

\theoremstyle{definition}

\theoremstyle{definition}

\begin{document}

\begin{titlepage}

\begin{center}
{\Large Consistent and robust inference in hazard probability and odds models with discrete-time survival data}

\vspace{.1in} Zhiqiang Tan\footnotemark[1]

\vspace{.1in}
\today
\end{center}

\footnotetext[1]{Department of Statistics, Rutgers University. Address: 110 Frelinghuysen Road,
Piscataway, NJ 08854. E-mail: ztan@stat.rutgers.edu.}

\paragraph{Abstract.}
For discrete-time survival data, conditional likelihood inference in Cox's hazard odds model is theoretically desirable but exact calculation
is numerical intractable with a moderate to large number of tied events. Unconditional maximum likelihood estimation over both regression coefficients
and baseline hazard probabilities can be problematic with a large number of time intervals.
We develop new methods and theory using numerically simple estimating functions, along with
model-based and model-robust variance estimation, in hazard probability and odds models.
For the probability hazard model, we derive as a consistent estimator the Breslow--Peto estimator, previously known
as an approximation to the conditional likelihood estimator in the hazard odds model.
For the odds hazard model, we propose a weighted Mantel--Haenszel estimator, which satisfies conditional unbiasedness given the numbers of events in addition
to the risk sets and covariates, similarly to the conditional likelihood estimator.
Our methods are expected to perform
satisfactorily in a broad range of settings, with small or large numbers of tied events corresponding to a large or small number of time intervals.
The methods are implemented in the R package \texttt{dSurvival}.


\end{titlepage}

\section{Introduction} \label{sect:intro}

Regression analysis with censored survival outcomes has been widely used and extensively studied. The subjects are covered in numerous articles and books
(e.g., Anderson et al.~1993; Cox \& Oaks 1984; Kalbfleisch \& Prentice 1980; Therneau \& Grambsch 2000).
The dominant approach is to use  Cox's (1972) proportional hazards models and conditional or partial  likelihood inference.
For continuous-time survival data, this approach is statistically desirable, where the baseline hazard function as an infinite-dimensional nuisance parameter
can be eliminated through conditional inference successively given the event times, and large sample theory can be developed using counting processes.
Moreover, this approach is computationally convenient, because the partial log-likelihood function is concave in a coefficient vector in the proportional hazard model.


Regression analysis with discrete-time survival data, however, has been developed to a limited extent, even though such data arise frequently in various applications (e.g., Allison 1982; Willett \& Singer 2004).
As remarked by Cox (1972), ``Unfortunately it is quite likely in applications that the data will be recorded in a form involving ties."
The number of tied events can be substantial, depending on the discrete-time units used to record the survival data.
There are broadly three types of existing methods for handling discrete-time survival data.
The first type is to use Cox's (1972) discrete-time version of proportional hazards models on hazard odds ratios and apply conditional inference given the numbers of events.
This method retains the statistical superiority of eliminating the baseline hazard probabilities as nuisance parameters,
but exact calculation is numerically intractable with a moderate or large number of ties.
The second type of methods employ various ad hoc approximations to conditional likelihood estimation (Breslow 1974; Efron 1977; Peto 1972).
These methods are often considered to yield satisfactory results with a small number of ties, but there remains the difficulty of handling a relatively large number of ties.
Statistical properties of these methods seem to be ambiguous.
In fact, the estimators of Breslow (1974) and Efron (1977) would in general be inconsistent under Cox's discrete-time proportional hazards model.
The third type of methods resort to unconditional maximum likelihood over both regression coefficients and baseline hazard probabilities,
either with pooled logistic regression corresponding to Cox's discrete-time model, or complementary log-log regression induced by
grouping observations under Cox's continuous-time model (Prentice \& Gloeckler 1978).
While such methods are appropriate for a small number of time intervals, statistical performance of maximum likelihood estimation can be problematic
in the presence of many time intervals, which leads to the same number of nuisance parameters.

We develop new methods and theory for regression analysis with discrete-time survival data, while accommodating a broad range of data configurations,
including a small number of time intervals and large numbers of tied event times, or a large number of time intervals and small numbers of tied event times.
In contrast with previous methods, we derive numerically simple estimating equations, motivated by but distinct from conditional or unconditional likelihood inference,
and study model-based and model-robust statistical properties in two classes of regression models. The first model deals with how hazard probability ratios are associated with covariates,
whereas the second model is Cox's discrete-time proportional hazards model on hazard odds ratios. \vspace{-.05in}
\begin{itemize} \addtolength{\itemsep}{-.1in}
\item We derive as a consistent estimator the Breslow--Peto estimator in the hazard probability model, even though the same estimator is known
as an approximation to the partial likelihood estimator in the hazard odds model.
We find that the model-based asymptotic variance is no greater than the limit of the commonly used model-based variance estimator
for the Breslow--Peto estimator.

\item We propose a weighted Mantel--Haenszel estimator in Cox's hazard odds model, such that it is numerically tractable and expected to achieve similar
performance as the conditional likelihood estimator. We show that
the weighted Mantel--Haenszel estimating function is conditionally unbiased given the numbers of events in addition to the risk sets and covariates,
similarly to the conditional likelihood estimator.

\item We study both model-based and model-robust variance estimation. As a useful complement to model-based inference, model-robust variance estimation
captures sampling variation of a point estimator with possible misspecification of a posited model. Moreover, the influence function obtained sheds light on
the reduction of the asymptotic variance if the model is correctly specified.
\end{itemize}
See White (1982) and Manski (1988) for asymptotic theory in misspecified models, and
Buja et al.~(2019) for a recent discussion on model-robust variance estimation.

An important, technical feature of our methods is that the sample estimating functions in regression coefficients are carefully constructed to achieve various unbiasedness properties, which are
relevant in different asymptotic settings. First, the population estimating functions, defined as the probability limits of the sample estimating functions as
the sizes of all risk sets increase to infinity, are unconditionally unbiased.
Moreover, the sample estimating functions are conditionally unbiased successively given the risk sets and covariates.
This property can be exploited to establish consistency of the point and variance estimators, while allowing some risk-set sizes bounded in probability as the sample size increases.
Finally, under the hazard odds model, the weighted Mantel--Haenszel estimating function is also conditionally unbiased given the numbers of events in addition to the risk sets and covariates.
Consistency of the point and variance estimators can be obtained while conditioning on the numbers of events.

\section{Data and models} \label{sec:data}

Suppose that survival data, possibly right-censored, and covariates are obtained as $\{(Y_i, \delta_i, X_i): i=1,\ldots,N\}$ from $N$ individuals,
where $Y_i = \min(T_i,C_i)$, $\delta_i = 1\{T_i \le C_i\}$, $T_i$ is an event time such as death time, $C_i$ is a censoring time, and $X_i$ is a covariate vector.
Assume that $\{(T_i,C_i,X_i): i=1,\ldots,N\}$ are independent and identically distributed copies of $(T,C,X)$, and hence
$\{(Y_i,\delta_i,X_i): i=1,\ldots,N\}$ are independent and identically distributed copies of $(Y,\delta,X)$ with $Y = \min(T,C)$ and $\delta = 1\{T \le C\}$.
In addition, assume that the censoring and event variables, $C$ and $T$, are independent conditionally on the covariate vector $X$.

In practice, survival data are usually recorded by grouping continuous or fine-scaled measurements, but such detailed data are not available to data analysis.
While discretization of an uncensored time into an interval is straightforward, there are different options
in assigning a censored time to a discrete value representing an interval.
See Kaplan \& Meier (1958), Thompson (1977), and Tan (2019, Supplement) for discussion on related issues.
Nevertheless, assume that there are discrete values, $0 = t_0 < t_1 < \cdots < t_J < t_{J+1}$, such that $(Y,\delta)$ and $(T,C)$ are properly
defined with $C \in \{t_0, t_1, \ldots, t_J\}$ and $T \in \{t_1, \ldots, t_J, t_{J+1}\}$ and the conditionally independent censoring assumption is satisfied.
An uncensored time in the interval $(t_{j-1},t_j]$ is encoded as $Y=t_j$ and $\delta=1$.
For the censored-early option,
a censored time in $[t_{j-1}, t_j)$ is encoded as $Y=t_{j-1}$ and $\delta=0$ and the observation is included in the risk set up to time $t_{j-1}$.
For the censored-late option,
a censored time in $[t_{j-1}, t_j)$ is encoded as $Y=t_j$ and $\delta=0$ and the observation is included in the risk set up to time $t_j$ (Cox 1972).

We study extensions of Cox's proportional hazards model to discrete-time survival data as described above.
For $j=1,\ldots,J$, the hazard probability at time $t_j$ given covariates $X=x$ is defined as
$\pi_j(x) = P( T = t_j | T \ge t_j, X=x)$.
This probability, under conditionally independent censoring, can be identified from observed data as
\begin{align*}
p_j( x) = P( Y =t_j, \delta=1 | Y \ge t_j, X=x) .
\end{align*}
The subset $\{Y \ge t_j\}$, called the risk set at time $t_j$, represents individuals who are event-free (or alive) just prior to time $t_j$.
In the following, we state probability and odds ratio models directly in terms of
the event probabilities $p_j(x)$, which coincide with the hazard probabilities $\pi_j(x)$ if conditionally independent censoring holds,
but otherwise remains empirically identifiable. For ease of interpretation, we treat $p_j(x)$ interchangeably with the hazard probabilities $\pi_j(x)$ whenever possible.

Consider two types of regression models on the hazard probabilities $p_j(x)$. The first places a parametric restriction on the probability ratios:
\begin{align}
p_j(x) = p_j(x_0) \me^{{(x-x_0)^\T \gamma^*}}, \quad j=1,\ldots,J, \label{eq:prob-model}
\end{align}
where $x_0$ is a fixed vector of covariates, for example $x_0=0$, $\gamma^*$ is an unknown coefficient vector, and
the baseline probabilities $\{p_j(x_0): j=1, \ldots, J\}$ are left to be unspecified. 
The second model places a parametric restriction on the odds ratios:
\begin{align}
\frac{p_j(x)}{1-p_j(x)} = \frac{p_j(x_0)}{1- p_j(x_0)} \me^{{(x-x_0)^\T \beta^*}}, \quad j=1,\ldots, J, \label{eq:odds-model}
\end{align}
where $\beta^*$ is an unknown coefficient vector. In the limit of arbitrarily small time intervals,
both models (\ref{eq:prob-model}) and (\ref{eq:odds-model}) can be seen to
reduce to a Cox proportional hazards model:
\begin{align*}
\lambda_t(x)= \lambda_t(x_0) \me^{{(x-x_0)^\T \alpha^*}}, 
\end{align*}
where the survival time $T$ is absolutely continuous with a hazard function $\lambda_t(x)$ given $X=x$, and $\alpha^*$ is an unknown coefficient vector.
However, for discrete survival data,  models (\ref{eq:prob-model}) and (\ref{eq:odds-model}) represent two alternative modeling approaches.

Model (\ref{eq:odds-model}) is known as the discrete-time version of Cox's (1972) propositional hazard model.
By comparison, model (\ref{eq:prob-model}) seems to be previously not studied, although it can also be called a proportional hazards model because
the hazard probability ratio $p_j(x)/p_j(x_0)$ is assumed to be constant in $j=1,\ldots,J$.
A potential limitation of model  (\ref{eq:prob-model}) is that the range of $p_j(x_0)$ or $p_j(x)$ as a probability between 0 and 1 may be violated for a fitted model,
especially if model (\ref{eq:prob-model}) is misspecified.
The chance of such violation can be small if model (\ref{eq:prob-model}) is correctly specified or approximately so.
Examination of fitted hazard probabilities can serve as diagnosis. See Section~\ref{sec:prob-model} for further discussion.

\section{Inference in hazard probability models} \label{sec:prob-model}

\noindent\textbf{Point estimation}.
To derive a point estimator for $\gamma^*$, we rewrite model (\ref{eq:prob-model}) as
\begin{align}
P ( Y=t_j,\delta=1 | Y\ge t_j, X=x) = \me^{\gamma^*_{0j} + x^\T \gamma^*}, \quad j=1,\ldots, J, \label{eq:prob-model2}
\end{align}
where $\gamma^*_0 = (\gamma^*_{01}, \ldots, \gamma^*_{0J})^\T$ is a vector of unknown intercepts and $\gamma^*$ is as before.
Our estimators for $(\gamma^*_0, \gamma^*)$ are defined jointly as a solution $(\hat\gamma_0, \hat\gamma)$ to
\begin{align}
& \sum_{i: Y_i \ge t_j} \left( D_{ji} -  \me^{\gamma_{0j} + X_i^\T \gamma} \right)  = 0 , \quad j=1,\ldots, J, \label{eq:gam-est-a}\\
& \sum_{j=1}^J \sum_{i: Y_i \ge t_j} \left( D_{ji} - \me^{\gamma_{0j}+X_i^\T \gamma} \right)  X_i = 0 , \label{eq:gam-est-b}
\end{align}
where $D_{ji} = 1\{ Y_i = t_j,\delta_i=1\}$, equal to 1 if $Y_i=t_j$ and $\delta_i=1$ or 0 otherwise.
Equation (\ref{eq:gam-est-a}) depends only on the data from $j$th risk set $\{i: Y_i \ge t_j\}$,
whereas equation (\ref{eq:gam-est-b}) involves the data combined from all $J$ risk sets.
Within the $j$th risk set, the associated estimating functions in $(\gamma_{0j}, \gamma)$ are $ \sum_{i: Y_i \ge t_j} (D_{ji} - \me^{\gamma_{0j} + X_i^\T \gamma}) (1,X^\T)^\T$,
corresponding to quasi-likelihood score functions in model (\ref{eq:prob-model2}), viewed as a conditional moment restriction model  with the Poisson logarithmic link for $D_{ji}$ given $X_i$.

Solving (\ref{eq:gam-est-a}) for $ \gamma_{0j}$ with fixed $\gamma$ and substituting into (\ref{eq:gam-est-b}) shows that
\begin{align}
\me^{\hat\gamma_{0j}} = \frac{\sum_{i: Y_i \ge t_j} D_{ji}} {\sum_{i: Y_i \ge t_j} \me^{ X_i^\T \hat\gamma}}. \label{eq:gam0-est}
\end{align}
and $\hat\gamma$ can be determined from the closed-form estimating equation
\begin{align}
\sum_{j=1}^J \sum_{i: Y_i \ge t_j} \left( D_{ji} - \frac{\sum_{l: Y_l \ge t_j} D_{jl}} {\sum_{l: Y_l \ge t_j} \me^{ X_l^\T \gamma}} \me^{X_i^\T \gamma} \right)  X_i = 0 . \label{eq:gam-est}
\end{align}
By an exchange of indices $i$ and $l$, equation (\ref{eq:gam-est}) can be equivalently written as
\begin{align}
\sum_{j=1}^J \sum_{i: Y_i \ge t_j}   D_{ji} \left( X_i - \frac{\sum_{l: Y_l \ge t_j} \me^{X_l^\T \gamma} X_l} {\sum_{l: Y_l \ge t_j} \me^{ X_l^\T \gamma}} \right) = 0 , \label{eq:bres-est}
\end{align}
which is originally the estimating equation satisfied by the Breslow's (1974) and Peto's (1972) modification of the partial likelihood estimator
to deal with tied event times in Cox's (continuous-time) proportional hazards model.
Hence the estimator $\hat\gamma$ can be referred to as the Breslow--Peto estimator.
Moreover, $\me^{\hat\gamma_{0j}}$ in (\ref{eq:gam0-est}) coincides with Breslow's (1974) estimator of the baseline hazard function.
As a result, the difference
$$  D_{ji} - \me^{ \hat\gamma_{0j} + X_i^\T \hat\gamma} =
D_{ji} - \frac{\sum_{l: Y_l \ge t_j} D_{jl}} {\sum_{l: Y_l \ge t_j} \me^{ X_l^\T \hat \gamma}} \me^{X_i^\T \hat\gamma} $$
evaluated at $(\hat\gamma_{0j},\hat\gamma)$ is the martingale residual of $i$th individual at time $t_j$ in Therneau et al. (1990), adapted to our setting of discrete survival data.

The probability ratio $p_j(x)/p_j(x_0)$ is generally closer to 1 than the odds ratios $p_j(x) (1-p_j(x_0)) /\{ p_j(x_0)(1-p_j(x))\}$.
Hence our derivation explains the observation that the Breslow--Peto approximation often produces a conservative bias in estimating
regression coefficients too close to 0 in proportional hazards models (Cox \& Oaks 1984).

\vspace{.1in}
\noindent\textbf{Model-robust inference}.
We study model-robust inference using $\hat\gamma$ with possible misspecification of model (\ref{eq:prob-model2}), similarly as in Lin \& Wei (1989)
for robust inference in Cox's proportional hazards model.
Denote  $R_{ji} = 1\{Y_i \ge t_j\}$, in addition to $D_{ji} = 1\{ Y_i = t_j,\delta_i=1\}$. Estimating equation (\ref{eq:gam-est}) can be written as
$\sum_{j=1}^J \hat \zeta_j(\gamma) = 0$, where
\begin{align*}
\hat \zeta_j (\gamma) = \frac{1}{n} \sum_{i=1}^n R_{ji} \left( D_{ji} - \frac{\sum_{l=1}^n R_{jl} D_{jl} } {\sum_{l=1}^n R_{jl} \me^{ X_l^\T \bar\gamma}} \me^{X_i^\T \bar \gamma} \right)  X_i .
\end{align*}
Under suitable regularity conditions, it can be shown that $\hat\gamma$ converges in probability to a target value $\bar\gamma$, defined as a unique solution to
the population version of (\ref{eq:gam-est}) or equivalently (\ref{eq:bres-est}):
\begin{align}
0 & = \sum_{j=1}^J E\left[ R_j \left\{ D_j - \frac{E( \tilde R_j\tilde D_j)} {E(\tilde R_j \me^{ \tilde X^\T \gamma})} \me^{X^\T \gamma} \right\} X \right] \label{eq:gam-est-pop} \\
& = \sum_{j=1}^J E \left[ R_j D_j \left\{ X - \frac{E( \tilde R_j \me^{\tilde X^\T \gamma} \tilde X )} {E(\tilde R_j \me^{ \tilde X^\T \gamma})}  \right\} \right], \label{eq:bres-est-pop}
\end{align}
where $R_j = 1\{Y \ge t_j\}$, $D_j = 1\{ Y = t_j,\delta=1\}$, and $(\tilde R_j, \tilde D_j, \tilde X)$ are defined from $(\tilde Y,\tilde \delta,\tilde X)$ identically distributed as $(Y,\delta,X)$.
Equivalently, $\bar\gamma$ is a unique maximizer of the objective function (which is concave in $\gamma$):
\begin{align}
 \sum_{j=1}^J E \left[ R_j D_j \left\{ X^\T \gamma - \log E( \tilde R_j \me^{ \tilde X^\T \gamma} ) \right\} \right] . \label{eq:bres-obj}
\end{align}
Moreover, $\hat\gamma$ can be shown to admit the asymptotic expansion
\begin{align}
\hat\gamma - \bar\gamma & = B(\bar\gamma)^{-1}
 \sum_{j=1}^J \hat\zeta_j(\bar\gamma) + o_p(n^{-1/2}), \label{eq:gam-expan}
\end{align}
where  $B(\gamma)$ is the negative Hessian of objective function (\ref{eq:bres-obj}), that is,
\begin{align*}
& B(\gamma) = \sum_{j=1}^J E(\tilde R_j \tilde D_j)
E \left[ \frac{R_j \me^{X^\T\gamma}}{E(\tilde R_j \me^{ \tilde X^\T \gamma}) } \left\{ X - \frac{E( \tilde R_j \me^{\tilde X^\T \gamma} \tilde X )} {E(\tilde R_j \me^{ \tilde X^\T \gamma})}  \right\}^{\otimes 2} \right].
\end{align*}
Throughout, $x^{\otimes 2} = x x^\T$ for a vector $x$.
From (\ref{eq:gam-expan}), the following result can be deduced, provided that the probability of survival beyond time $t_J$ (which is the largest possible value of the censoring variable)
is bounded away from 0.
This boundedness condition is standard in large sample theory for survival analysis (e.g., Anderson et al.~1993, Condition VII.2.1), although further investigation can be of interest.

\begin{pro} \label{pro:gam-model-robust}
Assume that 
$P(T > t_J) \ge p_0$ for a constant $p_0>0$.
Then  $n^{1/2} (\hat\gamma-\bar\gamma)$ converges in distribution to $\N(0, V )$ as $n\to \infty$, where $V = B(\bar\gamma)^{-1} A(\bar\gamma) B(\bar\gamma)^{-1} $,
$B(\gamma)$ is defined as above, $A(\gamma) = \var \{ \sum_{j=1}^J h_j( Y, \delta, X; \gamma) \}$, and
\begin{align*}
 h_j(Y, \delta, X;   \gamma) = R_j \left\{ D_j - \frac{E( \tilde R_j\tilde D_j)} {E(\tilde R_j \me^{ \tilde X^\T \gamma})} \me^{X^\T \gamma} \right\}
 \left\{ X - \frac{E( \tilde R_j \me^{\tilde X^\T \gamma} \tilde X )} {E(\tilde R_j \me^{ \tilde X^\T \gamma})}  \right\}.
\end{align*}
Moreover, a consistent estimator of $V$ is $\hat V_{\text{r}}=\hat B^{-1}(\hat\gamma) \hat A (\hat\gamma) \hat B^{-1} (\hat\gamma) $,
where
\begin{align*}
& \hat B(\gamma) = \frac{1}{n} \sum_{j=1}^J  \left( \sum_{l=1}^n R_{jl} D_{jl} \right)
 \sum_{i=1}^n\left[ \frac{R_{ji} \me^{X_i^\T\gamma}}{\sum_{l=1}^n R_{jl} \me^{ X_l^\T \gamma} } \left\{ X_i - \frac{\sum_{l=1}^n R_{jl} \me^{X_l^\T \gamma} X_l } {\sum_{l=1}^n
 R_{jl} \me^{ X_l^\T \gamma}}  \right\}^{\otimes 2} \right], \\
& \hat A (\gamma) =  \frac{1}{n} \sum_{i=1}^n \left\{ \sum_{j=1}^J \hat h_j( Y_i, \delta_i, X_i; \gamma) \right\}^{\otimes2},
\end{align*}
and $\hat h_j(Y,\delta,X; \gamma)$ is defined as $h_j(Y,\delta, X;\gamma)$ with
$E (\tilde R_j \tilde D_j)$, $E(\tilde R_j \me^{ \tilde X^\T \gamma})$, and $E( \tilde R_j \me^{\tilde X^\T \gamma} \tilde X )$ replaced by the
sample averages
$n^{-1} \sum_{i=1}^n R_{ji} D_{ji}$, $n^{-1} \sum_{i=1}^n R_{ji} \me^{ \tilde X_i^\T \gamma}$, and $n^{-1}\sum_{i=1}^n R_{ji} \me^{\tilde X_i^\T \gamma} X_i $.
\end{pro}

Proposition~\ref{pro:gam-model-robust} can be formally seen as an extension of Lin \& Wei's (1989) result on the partial likelihood estimator
in Cox's continuous-time model to the Breslow--Peto estimator used to handle tied event times.
Lin \& Wei's approach would use the Breslow--Peto modified score equation (\ref{eq:bres-est})  and  derive the asymptotic variance $V$ with
$h_j (Y,\delta,X;\gamma)$ defined as a correction to the modified score function
\begin{align*}
 h_j(Y, \delta, X; \gamma) = R_j D_j \left\{ X - \frac{E( \tilde R_j \me^{\tilde X^\T \gamma} \tilde X )} {E(\tilde R_j \me^{ \tilde X^\T \gamma})}  \right\}
 - \frac{E( \tilde R_j\tilde D_j)\me^{X^\T \gamma}   } {E(\tilde R_j \me^{ \tilde X^\T \gamma})}
 R_j \left\{ X - \frac{E( \tilde R_j \me^{\tilde X^\T \gamma} \tilde X )} {E(\tilde R_j \me^{ \tilde X^\T \gamma})}  \right\}.
\end{align*}
From our approach, the asymptotic variance $V$ can be equivalently derived using the estimating equation (\ref{eq:gam-est}) based on the conditional moment model (\ref{eq:prob-model2}), and hence
$h_j (Y,\delta,X;\gamma)$ is obtained as a correction to the associated estimating function
\begin{align*}
 h_j(Y, \delta, X; \gamma) = R_j \left\{ D_j - \frac{E( \tilde R_j\tilde D_j) \me^{X^\T \gamma} } {E(\tilde R_j \me^{ \tilde X^\T \gamma})}\right\} X
 - \frac{E( \tilde R_j \me^{\tilde X^\T \gamma} \tilde X )} {E(\tilde R_j \me^{ \tilde X^\T \gamma})}   R_j \left\{ D_j - \frac{E( \tilde R_j\tilde D_j) \me^{X^\T \gamma} } {E(\tilde R_j \me^{ \tilde X^\T \gamma})} \right\} .
\end{align*}
As shown in (\ref{eq:simple-A}), this representation is useful for simplification of the asymptotic variance $V$ if model (\ref{eq:prob-model2}) is correct.
By numerical evaluation, the variance estimator $\hat V_{\text{r}}$ also appears to coincide with the robust variance estimator for the Breslow--Peto estimator in the R package \texttt{survival},
although no justification was provided.

\vspace{.1in}
\noindent\textbf{Model-based inference}.
We study model-based inference using $\hat\gamma$ when model (\ref{eq:prob-model2}) is correctly specified.
Under this assumption, $\hat\gamma$ is a consistent estimator of $\gamma^*$, with $\bar\gamma = \gamma^*$ satisfying the population estimating equation (\ref{eq:gam-est-pop}):
\begin{align}
 & E\left[ R_j \left\{ D_j - \frac{E( \tilde R_j\tilde D_j)} {E(\tilde R_j \me^{ \tilde X^\T \gamma^*})} \me^{X^\T \gamma^*} \right\} X \right]
 = E\left[ R_j \left\{ D_j - \me^{\gamma^*_{0j}} \me^{X^\T \gamma^*} \right\} X \right] = 0, \label{eq:gam-unbiased}
\end{align}
because $ E \{R_j (D_j - \me^{\gamma^*_{0j} + X^\T \gamma^*}) | X\} =0$ and
$\me^{\gamma^*_{0j}} = E( \tilde R_j\tilde D_j) / E(\tilde R_j \me^{ \tilde X^\T \gamma^*} )$ by (\ref{eq:prob-model2}).
The true value $\gamma^*$ also satisfies the Breslow--Peto equation (\ref{eq:bres-est-pop}), by the equivalence between (\ref{eq:gam-est-pop}) and (\ref{eq:bres-est-pop}).
This finding seems new. Interestingly, consistency of the Breslow--Peto
estimator $\hat\gamma$ under model (\ref{eq:prob-model2}) is revealed more directly when defined through the new estimating equation (\ref{eq:gam-est}) than through
the usual equation (\ref{eq:bres-est}).

There is also an interesting implication on model-based variance estimation. Under model (\ref{eq:prob-model2}),
the difference $D_j -  \me^{\gamma^*_{0j} + X^\T \gamma^* }$ has mean 0 conditionally on $R_j=1$ and $X$,
and the individual terms $h_j(Y,\delta,X;\gamma^*)$, $j=1,\ldots,J$, are uncorrelated with each other.
Then the asymptotic variance $V = B^{-1}(\gamma^*) A(\gamma^*) B^{-1}(\gamma^*)$ can be simplified such that
\begin{align}
A(\gamma^*) &= \sum_{j=1}^J \var \{h_j( Y, \delta, X; \gamma^*) \} \label{eq:simple-A} \\
& = \sum_{j=1}^J E\left[ R_j p_j(X) (1-p_j(X))
 \left\{ X - \frac{E( \tilde R_j \me^{\tilde X^\T \gamma^*} \tilde X )} {E(\tilde R_j \me^{ \tilde X^\T \gamma^*})}  \right\}^{\otimes 2} \right], \nonumber
\end{align}
where $p_j(X) = \me^{\gamma^*_{0j} + X^\T \gamma^* } =
\{E( \tilde R_j\tilde D_j) /E(\tilde R_j \me^{ \tilde X^\T \gamma^*})\} \me^{X^\T \gamma^*}$.
A model-based estimator for the asymptotic variance $V$
is then $\hat V_{\text{b}}=\hat B^{-1}(\hat\gamma) \hat A_{\text{b}} (\hat\gamma) \hat B^{-1} (\hat\gamma) $ with
\begin{align*}
& \hat A_{\text{b}}(\gamma) = \frac{1}{n} \sum_{j=1}^J
 \sum_{i=1}^n\left[  R_{ji} \hat p_j(X_i;\gamma) (1- \hat p_j(X_i;\gamma))   \left\{ X_i - \frac{\sum_{l=1}^n R_{jl} \me^{ X_l^\T \gamma} X_l } {\sum_{l=1}^n
 R_{jl} \me^{ X_l^\T \hat\gamma}}  \right\}^{\otimes 2} \right],
\end{align*}
where  $\hat p_j(X;\gamma)  =
\{(\sum_{l=1}^n R_{jl} D_{jl}) / (\sum_{l=1}^n R_{jl} \me^{ X_l^\T \gamma})\} \me^{X^\T \gamma}$.
By direct comparison, $A(\gamma^*)$ and $\hat A_{\text{b}}(\hat\gamma)$  are no greater than respectively $B(\gamma^*)$ and $\hat B(\hat\gamma)$.

\begin{cor} \label{cor:gam-bres-var}
Suppose that model (\ref{eq:prob-model2}) is correctly specified. Then the asymptotic variance $V$ for $n^{1/2} (\hat\gamma-\gamma^*)$ is,
in the order on variance matrices, no greater than the  $B^{-1} (\gamma^*)$, and the variance estimator $\hat V_{\text{b}}$ is no greater
than $\hat B^{-1} (\hat\gamma)$, the commonly used variance estimator for the Breslow--Peto estimator $\hat\gamma$.
\end{cor}

To accommodate small risk sets, we outline asymptotic theory conditionally on the risk sets and covariates and propose an improved model-based variance estimator.
In fact, the foregoing justification of the asymptotic variance $V$ and the variance estimators $\hat V_{\text{r}}$ and $\hat V_{\text{b}}$ rely on
the assumption that all $J$ risk sets are sufficiently large to ensure convergence of
the sample averages $n^{-1}\sum_{l=1}^n R_{jl} D_{jl}$, $n^{-1}\sum_{l=1}^n R_{jl} \me^{ X_l^\T \gamma}$,
and $n^{-1} \sum_{l=1}^n R_{jl} \me^{ X_l^\T \gamma} X_l$ to their corresponding expectations for $j=1,\ldots,J$.
Alternatively,  asymptotic properties of $\hat\gamma$ can be studied by exploiting the conditional unbiasedness of individual terms of the sample estimating function in
(\ref{eq:gam-est}) under model (\ref{eq:prob-model2}):
\begin{align}
E \left\{ \hat \zeta_j(\gamma^*) | R_{j,1:n}, X_{1:n} \right\} = 0, \quad j=1,\ldots, J, \label{eq:gam-unbiased-mean}
\end{align}
where $R_{j,1:n} = (R_{j1},\ldots, R_{jn})$ and $X_{1:n}=(X_1, \ldots, X_n)$.
This is a more elaborate property than unconditional unbiasedness (\ref{eq:gam-unbiased}).
Under suitable regularity conditions similar as in fixed-design analysis of regression models,
it can be shown that if model (\ref{eq:prob-model2}) is correctly specified, then
 $n^{1/2} (\hat\gamma-\gamma^*)$ converges in distribution to $\N(0, V_2)$ as $n\to \infty$, where $V_2 = B_2(\gamma^*)^{-1} A_2(\gamma^*) B_2(\gamma^*)^{-1} $,
$B_2(\gamma) = \plim_{n\to\infty} \hat B(\gamma) $, $A_2(\gamma) = \plim_{n\to\infty} \sum_{j=1}^J v_j(\gamma)$, and
$v_j(\gamma) = n\, \var\{ \hat\zeta_j(\gamma) | R_{j,1:n}, X_{1:n} \}$, that is,
\begin{align*}
v_j(\gamma) = \frac{1}{n} \var
\left\{ \sum_{i=1}^n R_{ji} \left( D_{ji} - \frac{\sum_{l=1}^n R_{jl} D_{jl} } {\sum_{l=1}^n R_{jl} \me^{ X_l^\T \gamma}} \me^{X_i^\T \gamma} \right)  X_i \Big| R_{j,1:n}, X_{1:n}\right\} .
\end{align*}
In the case where $P(T>t_J)$ is bounded away from 0 and all $J$ risk sets are of sizes increasing to $\infty$,
the asymptotic variance $V_2$ reduces to $V$ in Proposition~\ref{pro:gam-model-robust}.

For the asymptotic variance $V_2$, our proposed estimator is $\hat V_{\text{b2}} = \hat B(\hat\gamma)^{-1} \hat A_{\text{b2}}(\hat \gamma) \hat B(\hat\gamma)^{-1} $,
where $\hat B(\gamma)$ is as in Proposition~\ref{pro:gam-model-robust}, $\hat A_{\text{b2}}(\gamma) = \sum_{j=1}^J \{\hat v_j(\gamma) + \hat v_j^\T(\gamma)\}/2$, and
\begin{align*}
\hat v_j(\gamma) = \frac{1}{n} \sum_{i=1}^n R_{ji} (1-D_{ji}) \me^{X_i^\T \gamma} \frac{\sum_{l=1}^n R_{jl} \me^{X_l^\T \gamma} (X_i-X_l) \sum_{k=1}^n R_{jk} D_{jk} (X_i-X_k)^\T }
{(\sum_{l=1}^n R_{jl} \me^{X_l^\T \gamma} )^2} .
\end{align*}
The matrix $\hat v_j(\gamma)$ is in general not symmetric, and $\{\hat v_j(\gamma) + \hat v_j^\T(\gamma)\}/2$ serves as a symmetrized version.
The following properties can be established.

\begin{pro} \label{pro:gam-model-based}
(i) Suppose that model (\ref{eq:prob-model2}) is correctly specified.  For $j=1,\ldots, J$, $\hat v_j(\gamma^*) $ is a conditionally unbiased estimator for $v_j(\gamma^*)$, that is,
\begin{align*}
E \{ \hat v_j(\gamma^*)  | R_{j,1:n}, X_{1:n} \}= v_j(\gamma^*).
\end{align*}
Hence $\hat V_{\text{b2}} $ can be a consistent estimator for $V_2$ even if some risk sets are of sizes which are bounded in probability as  $J\to\infty$ and $n\to\infty$.\\
(ii) Suppose that at most one event is observed in each risk set $\{i: R_{ji}=1\}$ for $j=1,\ldots,J$. Then
$\hat A_{\text{b2}}(\gamma)$ is identical to $\hat B(\gamma)$ and the variance estimator $\hat V_{\text{b2}}$ is identical to $\hat B(\hat\gamma)^{-1}$,
the usual variance estimator for the partial likelihood estimator (i.e., the Breslow--Peto estimator in the absence of tied events).
\end{pro}

Property (ii) in Proposition~\ref{pro:gam-model-based} shows that the variance estimator $\hat V_{\text{b2}}$ is the same as $\hat B^{-1}(\hat\gamma)$ in the extreme case where
there are no tied events and the Breslow--Peto estimator reduces to the maximum partial likelihood estimator.
In contrast, the variance estimator $\hat V_{\text{b}}$ in this case remains smaller than $\hat B^{-1}(\hat\gamma)$.

The variance estimator $\hat V_{\text{b2}}$ is an extension of a model-based variance estimator  in Tan (2019) for the Breslow--Peto estimator in a probability ratio model
for analysis of $2\times 2$ tables and two-sample survival analysis. See the Supplement for details of the relationship.
For $2\times 2$ tables, the variance estimator is designed to
be consistent in two asymptotic settings, either with a fixed number of large tables or with a large number of possibly sparse tables.
These two settings are originally considered for Mantel--Haenszel estimation of common odds ratios in $2\times 2$ tables (Robins et al.~1986).

\vspace{.1in}
\noindent\textbf{Estimation of survival probabilities}.
We discuss estimation of survival probabilities for individuals with fixed covariates $x_0$. For simplicity, assume that $x_0=0$ in model (\ref{eq:prob-model});
otherwise the covariates can be recentered. Then the hazard probability $p_j(x_0)$ is identified as $\me^{\gamma_{0j}}$, and can be estimated as $\hat p_j(x_0)=\me^{\hat\gamma_{0j}}$ by (\ref{eq:gam0-est}).
The $k$th survival probability, defined as $P_k(x_0) = P( T > t_k |X=x_0)$, can be estimated as
\begin{align}
\hat P_k( x_0) = \prod_{j=1}^k \{ 1- \hat p_j(x_0) \} = \prod_{j=1}^k \left( 1- \me^{\hat\gamma_{0j}} \right) , \quad k=1,\ldots, J. \label{eq:gam-surv}
\end{align}
This is a discrete version of the product-limit estimator of the baseline survival function. Unless all $\hat p_j(x_0)$ are sufficiently small,
the estimator (\ref{eq:gam-surv}) is distinct from an alternative estimator, $\me^{-\sum_{j=1}^k \me^{\hat\gamma_{0j}}}$, where $\sum_{j=1}^k \me^{\hat\gamma_{0j}}$ is called the cumulative hazard.
The alternative estimator is often used with continuous-time data in the R package \texttt{survival}.
A potential disadvantage is that the estimator (\ref{eq:gam0-est}) for the hazard probability and hence (\ref{eq:gam-surv}) for the survival probability
may be negative, in general due to the fact that the right hand side of model (\ref{eq:prob-model2}) is not restricted to be no greater than 1.
Such negative estimates may also occur due to estimation error, particularly in the right tail.

The standard errors for $\hat P_k(x_0)$ can be obtained using Taylor expansions (or the delta method) and either model-robust or model-based variance estimator for $\hat\gamma$.
In particular, model-robust variance estimation for $\hat P_k(x_0)$ involves use of the influence function of $\hat\gamma$ depending on the data from all $J$ risk sets.
Model-based variance estimation for $\hat P_k(x_0)$ admits a decomposition similar to variance estimation of the cumulative hazard in Tsiatis (1981).
See the Supplement for detailed derivation and formulas.

\section{Inference in hazard odds models} \label{sec:odds-model}

\noindent\textbf{Point estimation}.
To derive a point estimator for $\beta^*$, we rewrite model (\ref{eq:odds-model}) as
\begin{align}
P ( Y=t_j,\delta=1 | Y\ge t_j, X=x) = \expit(\beta^*_{0j} + x^\T \beta^*), \quad j=1,\ldots, J, \label{eq:odds-model2}
\end{align}
where $\expit(c) = \me^c/(1+\me^c)$, $\beta^*_0 = (\beta^*_{01}, \ldots, \beta^*_{0J})^\T$ is a vector of unknown intercepts and $\beta^*$ is as before.
Our estimators for $(\beta^*_0, \beta^*)$ are defined jointly as a solution $(\hat\beta_0, \hat\beta)$ to
\begin{align}
& \sum_{i: Y_i \ge t_j} \left\{ D_{ji} - (1-D_{ji})  \me^{\beta_{0j} + X_i^\T \beta} \right\} = 0 , \quad j=1,\ldots, J, \label{eq:bet-est-a}\\
& \sum_{j=1}^J \frac{\sum_{l: Y_l\ge t_j} (1-D_{jl}) \me^{X_l^\T \beta}}{\sum_{l: Y_l\ge t_j} \me^{X_l^\T \beta}} \sum_{i: Y_i \ge t_j}
\left\{ D_{ji} - (1-D_{ji}) \me^{\beta_{0j}+X_i^\T \beta} \right\}  X_i = 0 , \label{eq:bet-est-b}
\end{align}
where $D_{ji} = 1\{ Y_i = t_j,\delta_i=1\}$ as in Section~\ref{sec:prob-model}.
Similarly as (\ref{eq:gam-est-a})--(\ref{eq:gam-est-b}), equation (\ref{eq:bet-est-a}) depends only on the data from $j$th risk set $\{i: Y_i \ge t_j\}$,
whereas equation (\ref{eq:bet-est-b}) involves the data combined from all $J$ risk sets.
Within the $j$th risk set, the associated estimating functions in $(\beta_{0j}, \beta)$ are
\begin{align*}
& \sum_{i: Y_i \ge t_j} \{D_{ji} - (1-D_{ji})\me^{\beta_{0j} + X_i^\T \beta}\} (1,X^\T)^\T \\
& = \sum_{i: Y_i \ge t_j}    \left\{ 1- \frac{1-D_{ji}}{\expit(-\beta_{0j}- X_i^\T \beta)} \right\} (1,X^\T)^\T
\end{align*}
which, interestingly, corresponds to the estimating functions for calibrated estimation (Tan 2020a) in logistic regression model (\ref{eq:odds-model2}) for $1-D_{ji}$ given $X_i$.
The $j$-dependent factor $\{\sum_{l: Y_l\ge t_j} (1-D_{jl}) \me^{X_l^\T \beta}\}/(\sum_{l: Y_l\ge t_j} \me^{X_l^\T \beta})$ in (\ref{eq:bet-est-b}) is introduced to
achieve reduction, as discussed below, to the weighted Mantel--Haenszel estimator in two-sample survival analysis in Tan (2019)
and to the maximum partial likelihood estimator in the case of only one event per risk set in Cox's continuous-time model.
In addition, use of this factor is crucial for achieving conditional unbiasedness as in (\ref{eq:bet-unbiased-mean}) and (\ref{eq:bet-unbiased-mean2}).

Solving (\ref{eq:bet-est-a}) for $ \beta_{0j}$ with fixed $\beta$ and substituting into (\ref{eq:bet-est-b}) shows that
\begin{align}
\me^{\hat\beta_{0j}} = \frac{\sum_{i: Y_i \ge t_j} D_{ji}} {\sum_{i: Y_i \ge t_j} (1-D_{ji}) \me^{ X_i^\T \hat\beta}}. \label{eq:bet0-est}
\end{align}
and $\hat\beta$ can be determined from the closed-form estimating equation
\begin{align}
\sum_{j=1}^J \sum_{i: Y_i \ge t_j} \frac{ D_{ji} \sum_{l: Y_l \ge t_j} (1-D_{jl})\me^{ X_l^\T \beta} - (1-D_{ji}) \me^{X_i^\T \beta}\sum_{l: Y_l \ge t_j} D_{jl}}
{ \sum_{l: Y_l \ge t_j}\me^{ X_l^\T \beta} }  X_i = 0 . \label{eq:bet-est}
\end{align}
By some rearrangement, equation (\ref{eq:bet-est}) can be equivalently written as
\begin{align}
\sum_{j=1}^J \sum_{i: Y_i \ge t_j} D_{ji} \frac{ \sum_{l: Y_l \ge t_j} (1-D_{jl})\me^{ X_l^\T \beta} (X_i-X_l) }
{ \sum_{l: Y_l \ge t_j}\me^{ X_l^\T \beta} } =0 , \label{eq:cmh-est}
\end{align}
which closely resembles the Breslow--Peto estimating equation (\ref{eq:bres-est})
with only the additional factor $1-D_{jl}$  in front of $\me^{X_i^\T \beta}(X_i-X_l)$.
In fact, in the extreme case of only one event observed in each risk set $\{i: Y_i \ge t_j \}$ for $j=1,\ldots,J$, then
equation (\ref{eq:cmh-est}) is easily shown to be equivalent to (\ref{eq:bres-est}),
and hence the estimator $\hat\beta$ numerically coincides with the Breslow--Peto  or the maximum partial likelihood estimator $\hat\gamma$.

For two-sample survival analysis with a binary covariate $X$, estimating equation (\ref{eq:bet-est}) or (\ref{eq:cmh-est})
can be shown to yield the weighted Mantel--Haenszel estimator proposed in Tan (2019) as an extension of
Cochran's (1954) and Mantel \& Haenszel's (1959) estimation of common odds ratios in analysis of $2\times 2$ tables.
See the Supplement for details.
Hence the estimator $\hat\beta$ can also be called a weighted Mantel--Haenszel estimator.

\vspace{.1in}
\noindent\textbf{Model-robust inference}.
We study model-robust inference using $\hat\beta$ with possible misspecification of model (\ref{eq:odds-model2}), similarly as in Section~\ref{sec:prob-model}
for robust inference using $\hat\gamma$ in the hazard probability model.
Denote, as before, $R_{ji} = 1\{Y_i \ge t_j\}$ and $D_{ji} = 1\{ Y_i = t_j,\delta_i=1\}$.
Estimating equation (\ref{eq:bet-est}) can be written as $\sum_{j=1}^J \hat\tau_j (\beta)=0$, where
\begin{align*}
\hat \tau_j(\beta) = \frac{1}{n} \sum_{i=1}^n R_{ji} \frac{ D_{ji} \sum_{l=1}^n R_{jl} (1-D_{jl})\me^{ X_l^\T \beta} - (1-D_{ji}) \me^{X_i^\T \beta}\sum_{l=1}^n R_{jl} D_{jl}}
{ \sum_{l=1}^n R_{jl} \me^{ X_l^\T \beta} }  X_i .
\end{align*}
Under suitable regularity conditions, it can be shown that $\hat\beta$ converges in probability to a target value $\bar\beta$, defined as a unique solution to
the population version of (\ref{eq:bet-est}) or equivalently (\ref{eq:cmh-est}):
\begin{align}
0 & = \sum_{j=1}^J E\left[ R_j  \frac{ D_j E \{ \tilde R_j (1-\tilde D_j) \me^{ \tilde X^\T \beta} \} - (1-D_j) \me^{X^\T \beta} E( \tilde R_j \tilde D_j) }
{E(\tilde R_j \me^{ \tilde X^\T \beta})} X \right] \label{eq:bet-est-pop} \\
& = \sum_{j=1}^J E \left[ R_j D_j \frac{E\{ \tilde R_j (1-\tilde D_j) \me^{\tilde X^\T \beta}\} X -E\{ \tilde R_j (1-\tilde D_j)\me^{\tilde X^\T \beta}\tilde X \}}
{E(\tilde R_j \me^{ \tilde X^\T \beta})}  \right], \label{eq:cmh-est-pop}
\end{align}
where, as in Section~\ref{sec:odds-model},
$R_j = 1\{Y \ge t_j\}$, $D_j = 1\{ Y = t_j,\delta=1\}$, and $(\tilde R_j, \tilde D_j, \tilde X)$ are defined from $(\tilde Y,\tilde \delta,\tilde X)$ identically distributed as $(Y,\delta,X)$.
Moreover, $\hat\beta$ can be shown to admit the asymptotic expansion
\begin{align}
& \hat\beta - \bar\beta = H(\bar\beta)^{-1}
\sum_{j=1}^J \hat\tau_j(\bar\beta) + o_p(n^{-1/2}), \label{eq:bet-expan}
\end{align}
where $H(\beta)$ is the negative derivative matrix in $\beta^\T$ of
the right hand side of (\ref{eq:bet-est-pop}) or equivalently (\ref{eq:cmh-est-pop}), that is,
\begin{align*}
& H (\beta) = \sum_{j=1}^J  E\left[ \frac{R_j (1-D_j)  \me^{X^\T \beta}}{E(\tilde R_j \me^{ \tilde X^\T \beta})}  \left\{ E(\tilde R_j \tilde D_j)X-E(\tilde R_j \tilde D_j\tilde X) \right\}
\left\{ X^\T - \frac{E(\tilde R_j \me^{ \tilde X^\T \beta} \tilde X^\T) }{ E(\tilde R_j \me^{ \tilde X^\T \beta})} \right\} \right] .
\end{align*}
See the Supplement for details.
The matrix $H(\beta)$ is in general not symmetric, and hence cannot be an Hessian of a scalar objective function.
From (\ref{eq:bet-expan}), the following result can be deduced, provided that the probability of survival beyond time $t_J$ (which is the largest possible value of the censoring variable)
is bounded away from 0.

\begin{pro} \label{pro:bet-model-robust}
Assume that 
$P(T > t_J) \ge p_0$ for a constant $p_0>0$.
Then  $n^{1/2} (\hat\beta-\bar\beta)$ converges in distribution to $\N(0, \Sigma )$ as $n\to \infty$, where $\Sigma = H(\bar\beta)^{-1} G(\bar\beta) H(\bar\beta)^{\T^{-1}} $,
$H(\beta)$ is defined as above, $G(\beta) = \var \{ \sum_{j=1}^J g_j( Y, \delta, X; \beta) \}$, and
\begin{align*}
& g_j(Y, \delta, X;   \beta) \\
& = R_j \frac{D_j E(\tilde R_j (1-\tilde D_j)\me^{ \tilde X^\T \beta}) -  (1-D_j)\me^{X^\T \beta} E( \tilde R_j\tilde D_j)} {E(\tilde R_j \me^{ \tilde X^\T \beta})}
\left\{ X- \frac{ E( \tilde R_j (1-\tilde D_j) \me^{\tilde X^\T \beta}\tilde X)} {E( \tilde R_j (1-\tilde D_j) \me^{\tilde X^\T \beta})} \right\} \\
& \quad -  \frac{E(\tilde R_j \tilde D_j \tilde X) E(\tilde R_j (1-\tilde D_j)\me^{ \tilde X^\T \beta}) -  E(\tilde R_j(1-\tilde D_j)\me^{\tilde X^\T \beta}\tilde X) E( \tilde R_j\tilde D_j)}
{ E(\tilde R_j \me^{ \tilde X^\T \beta}) } \\
& \qquad \times
\left\{ \frac{ R_j  \me^{ X^\T \beta}}{E(\tilde R_j \me^{ \tilde X^\T \beta})} - \frac{ R_j (1-D_j) \me^{X^\T \beta} }{E( \tilde R_j (1-\tilde D_j) \me^{\tilde X^\T \beta})} \right\},
\end{align*}
which is denoted as $g_{j1} (Y,\delta,X;\beta) + g_{j2} (Y, \delta, X;\beta)$.
Moreover, a consistent estimator of $\Sigma$ is $\hat \Sigma_{\text{r}}=\hat H^{-1}(\hat\beta) \hat G (\hat\beta) \hat H^{\T^{-1}} (\hat\beta) $,
where
\begin{align*}
& \hat H(\beta) = \frac{1}{n}\sum_{j=1}^J  \sum_{i=1}^n \left[ \frac{R_{ji} (1-D_{ji})  \me^{X_i^\T \beta}} {\sum_{l=1}^n  R_{jl} \me^{ X_l^\T \beta} }  \left\{\sum_{l=1}^n R_{jl} D_{jl} (X_i-X_l) \right\}
\left\{ X_i^\T - \frac{\sum_{l=1}^n  R_{jl} \me^{ X_l^\T \beta} X_l^\T }{\sum_{l=1}^n R_{jl} \me^{ X_l^\T \beta} } \right\} \right] , \\
& \hat G (\beta) =  \frac{1}{n} \sum_{i=1}^n \left\{ \sum_{j=1}^J \hat g_j( Y_i, \delta_i, X_i; \beta) \right\}^{\otimes2},
\end{align*}
and $\hat g_j(Y,\delta,X; \beta)$ is defined as $g_j(Y,\delta, X;\beta)$ with all expectations replaced by the corresponding sample averages.
\end{pro}

From Proposition~\ref{pro:bet-model-robust}, the influence function of $\hat\beta$ is $H^{-1} (\bar\beta) \sum_{j=1}^J g_j( Y, \delta, X; \bar\beta)$.
Here $g_j(Y, \delta, X; \beta)$ consists of two terms. The first term, $g_{j1} (Y,\delta,X;\beta)$,
can be seen as a correction to the $j$th population estimating function in (\ref{eq:bet-est-pop}),
to account for the variation in substituting the estimator $\me^{\hat\beta_{0j}}$ for
$E( \tilde R_j \tilde D_j)  /E \{ \tilde R_j (1-\tilde D_j) \me^{ \tilde X^\T \beta} \}$ in the sample estimating equation (\ref{eq:bet-est}).
The second term, $g_{j2} (Y,\delta,X;\beta)$, is involved to further account for
substituting the factor $\{\sum_{l=1}^n R_{jl} (1-D_{jl})  \me^{ X_l^\T \beta} \} /( \sum_{l=1}^n R_{jl}  \me^{ X_l^\T \beta}) $
for the corresponding population quantity.
A similar interpretation of $g_j(Y, \delta, X; \beta)$ can also be obtained as a correction to the $j$th population estimating function in (\ref{eq:cmh-est-pop}).

\vspace{.1in}
\noindent\textbf{Model-based inference}.
We study model-based inference using $\hat\beta$ when model (\ref{eq:odds-model2}) is correctly specified.
Under this assumption, $\hat\beta$ is a consistent estimator of $\beta^*$, with $\bar\beta = \beta^*$ satisfying the population estimating equation (\ref{eq:bet-est-pop}):
\begin{align}
E\left[ R_j  \frac{ D_j E \{ \tilde R_j (1-\tilde D_j) \me^{ \tilde X^\T \beta^*} \} - (1-D_j) \me^{X^\T \beta^*} E( \tilde R_j \tilde D_j) }
{E(\tilde R_j \me^{ \tilde X^\T \beta^*})} X \right] = 0 , \label{eq:bet-unbiased}
\end{align}
because $ E [R_j \{D_j - (1-D_j)\me^{\beta^*_{0j} + X^\T \beta^*}\} | X] =0$ and
$\me^{\beta^*_{0j}} = E( \tilde R_j\tilde D_j) / E\{\tilde R_j (1-\tilde D_j)\me^{ \tilde X^\T \beta^*} \}$ by (\ref{eq:odds-model2}).
Equivalently, the true value $\beta^*$ also satisfies equation (\ref{eq:cmh-est-pop}).

Considerable simplification can be obtained for the model-based asymptotic variance for $\hat\beta$ in Proposition~\ref{pro:bet-model-robust}. Under model (\ref{eq:odds-model2}),
$g_j(Y,\delta,X; \beta^*)$ reduces to $g_{j1}(Y,\delta,X; \beta^*)$ only, because
  $g_{j2}(Y,\delta,X; \beta^*) \equiv 0$ due to (\ref{eq:bet-unbiased}).
Moreover, the difference $D_j - (1-D_j) \me^{\beta^*_{0j} + X^\T \beta^* }$ has mean 0 conditionally on $R_j=1$ and $X$,
and the individual terms $g_j(Y,\delta,X;\beta^*)$, $j=1,\ldots,J$, are uncorrelated with each other.
Then the asymptotic variance $\Sigma = H^{-1}(\beta^*) G(\beta^*) H^{\T^{-1}}(\beta^*)$ can be calculated such that
\begin{align}
G(\beta^*) &= \sum_{j=1}^J \var \{g_{j1}( Y, \delta, X; \beta^*) \} \label{eq:simple-G} \\
& = \sum_{j=1}^J E\left[ R_j \me^{\beta^*_{0j}+X^\T \beta^*} \frac{E^2(\tilde R_j(1-\tilde D_j) \me^{ \tilde X^\T \beta^*})} {E^2(\tilde R_j \me^{ \tilde X^\T \beta^*}) }
 \left\{ X - \frac{E( \tilde R_j (1-\tilde D_j)\me^{\tilde X^\T \beta^*} \tilde X )} {E(\tilde R_j (1-\tilde D_j)\me^{ \tilde X^\T \beta^*})}  \right\}^{\otimes 2} \right]. \nonumber
\end{align}
See the Supplement for details.
A model-based estimator for the asymptotic variance $\Sigma$
is then $\hat \Sigma_{\text{b}}=\hat H^{-1}(\hat\beta) \hat G_{\text{b}} (\hat\beta) \hat H^{-1} (\hat\beta) $, where $\hat G_{\text{b}}(\beta)$ is defined as
\begin{align*}
& \frac{1}{n} \sum_{j=1}^J \sum_{i=1}^n\left[ R_{ji} \me^{\beta_{0j}+X_i^\T \beta} \frac{(\sum_{l=1}^n R_{jl}(1- D_{jl}) \me^{X_l^\T \beta})^2 } {(\sum_{l=1}^n R_{jl} \me^{ X_l^\T \beta})^2 }
 \left\{ X_i - \frac{\sum_{l=1}^n R_{jl} (1-D_{jl}) \me^{ X_l^\T \beta} X_l } {\sum_{l=1}^n
 R_{jl} (1-D_{jl}) \me^{ X_l^\T \beta}}  \right\}^{\otimes 2} \right] ,
\end{align*}
with $\me^{\beta_{0j}}$ set to $(\sum_{l=1}^n R_{jl} D_{jl})/ \{\sum_{l=1}^n R_{jl} (1-D_{jl})\me^{ X_l^\T \beta}\} $.
The matrix $\hat G_{\text{b}}(\beta)$ is algebraically similar to the sample Hessian $\hat B(\gamma)$ in Section~\ref{sec:prob-model},
with only the additional factor $1-D_{jl}$ in front of  $\me^{ X_l^\T \beta}$ in various places.

Similarly as in Section~\ref{sec:prob-model}, we outline asymptotic theory conditionally on the risk sets and covariates and
propose an improved model-based variance estimator.
To accommodate small risk sets, asymptotic properties of $\hat\beta$ can be studied by exploiting the conditional unbiasedness of individual terms of the sample estimating function in
(\ref{eq:bet-est}) under model (\ref{eq:odds-model2}):
\begin{align}
E \left\{ \hat \tau_j(\beta^*) | R_{j,1:n}, X_{1:n} \right\} = 0, \quad j=1,\ldots, J, \label{eq:bet-unbiased-mean}
\end{align}
where $R_{j,1:n} = (R_{j1},\ldots, R_{jn})$ and $X_{1:n}=(X_1, \ldots, X_n)$.
This is a more elaborate property than unconditional unbiasedness (\ref{eq:bet-unbiased}).
Under suitable regularity conditions similar as in fixed-design analysis of regression models,
it can be shown that if model (\ref{eq:odds-model2}) is correctly specified, then
 $n^{1/2} (\hat\beta-\beta^*)$ converges in distribution to $\N(0, \Sigma_2)$ as $n\to \infty$, where $\Sigma_2 = H_2(\beta^*)^{-1} G_2(\beta^*) H_2(\beta^*)^{-1} $,
$H_2(\beta) = \plim_{n\to\infty} \hat H(\beta) $, $G_2(\beta) = \plim_{n\to\infty} \sum_{j=1}^J \sigma_j(\beta)$, and
$\sigma_j(\beta) = n\, \var\{ \hat\tau_j(\beta) | R_{j,1:n}, X_{1:n} \}$, that is,
\begin{align*}
\frac{1}{n} \var
\left\{ \sum_{i=1}^n R_{ji} \frac{ D_{ji} \sum_{l=1}^n R_{jl} (1-D_{jl})\me^{ X_l^\T \beta} - (1-D_{ji}) \me^{X_i^\T \beta}\sum_{l=1}^n R_{jl} D_{jl}}
{ \sum_{l=1}^n R_{jl} \me^{ X_l^\T \beta} }  X_i \Big| R_{j,1:n}, X_{1:n}\right\} .
\end{align*}
In the case where $P(T>t_J)$ is bounded away from 0 and all $J$ risk sets are of sizes increasing to $\infty$,
the asymptotic variance $\Sigma_2$ reduces to $\Sigma$ in Proposition~\ref{pro:bet-model-robust}.

For the asymptotic variance $\Sigma_2$, our proposed estimator is $\hat \Sigma_{\text{b2}} = \hat H^{-1}(\hat\beta) \hat G_{\text{b2}}(\hat \beta) \hat H^{-1} (\hat\beta)$,
where $\hat H(\beta)$ is as in Proposition~\ref{pro:bet-model-robust}, $\hat G_{\text{b2}}(\beta) = \sum_{j=1}^J \{\hat \sigma_j(\beta) + \hat \sigma_j^\T(\beta) \}/2$, and
\begin{align}
\hat \sigma_j(\beta) & = \frac{1}{n} \sum_{i=1}^n \left\{ R_{ji} (1-D_{ji}) \me^{X_i^\T \beta}  \frac{\sum_{l=1}^n R_{jl} D_{jl} \me^{X_l^\T \beta} (X_i-X_l)^{\otimes 2}}
{(\sum_{l=1}^n R_{jl} \me^{X_l^\T \beta} )^2} \right. \nonumber \\
& \quad \left. + R_{ji} \me^{X_i^\T \beta}  \frac{ \sum_{l=1}^n R_{jl} (1-D_{jl}) \me^{X_l^\T \beta} (X_i-X_l) \sum_{k=1}^n R_{jk} D_{jk} (X_i-X_k)^\T }
{(\sum_{l=1}^n R_{jl} \me^{X_l^\T \beta} )^2} \right\}. \label{eq:sigma-new}
\end{align}
The matrix $\hat \sigma_j(\beta)$ is in general not symmetric, and $\{\hat \sigma_j(\beta) + \hat \sigma_j^\T(\beta)\}/2$ serves as a symmetrized version.
The following properties can be established.

\begin{pro} \label{pro:bet-model-based}
(i) Suppose that model (\ref{eq:odds-model2}) is correctly specified. For $j=1,\ldots, J$, $\hat \sigma_j(\beta^*) $ is conditionally unbiased for $\sigma_j(\beta^*)$, that is,
\begin{align}
E \{ \hat \sigma_j(\beta^*)  | R_{j,1:n}, X_{1:n} \}= \sigma_j(\beta^*). \label{eq:bet-unbiased-var}
\end{align}
Hence $\hat \Sigma_{\text{b2}} $ can be a consistent estimator for $\Sigma_2$ even if some risk sets are of sizes which are bounded in probability as $J\to\infty$ and $n\to\infty$.\\
(ii) Suppose that at most one event is observed in each risk set $\{i: R_{ji}=1\}$ for $j=1,\ldots,J$.
Then $\hat\beta$ is identical to the maximum partial likelihood estimator, and $\hat H (\beta)$ and $\hat G_{\text{b2}}(\beta)$
are both identical to $\hat B(\beta)$. Hence
$\hat \Sigma_{\text{b2}}$ is identical to $\hat B^{-1} (\hat\beta)$,
the usual variance estimator for the maximum partial likelihood estimator.
\end{pro}

The variance estimator $\hat\sigma_j (\hat\beta)$ and the resulting sandwich variance $\hat \Sigma_{\text{b2}}$ represent a new development beyond
model-based variance estimation in Tan (2019) for the weighted Mantel--Haenszel estimator in an odds ratio model for analysis of $2\times 2$ tables and two-sample survival analysis.
The model-based variance estimator in Tan (2019) is adapted from that in Robins et al.~(1986) for the Mantel--Haenszel estimator of a common odds ratio in $2\times 2$ tables,
such that the variance estimator is consistent in both asymptotic settings of large tables and many sparse tables.
For two-sample analysis, the proposed estimator $\hat \Sigma_{\text{b2}}$ reduces to a variance estimator distinct from that in Robins et al.~(1986) as well as in Flander (1985).
See the Supplement for details.

For comparison, a suitable extension of model-based variance estimation from Robins et al.~(1986) and Tan (2019) to regression models is
$\hat \Sigma_{\text{b3}} = \hat H^{-1}(\hat\beta) \hat G_{\text{b3}}(\hat \beta) \hat H^{-1} (\hat\beta)$,
where $\hat H(\beta)$ is as in Proposition~\ref{pro:bet-model-robust}, $\hat G_{\text{b3}}(\beta) = \sum_{j=1}^J \tilde\sigma_j(\beta) $, and
\begin{align}
& \tilde \sigma_j(\beta) = \frac{1}{n} \sum_{i=1}^n R_{ji} (1-D_{ji}) \me^{X_i^\T \beta} \times \nonumber \\
& \quad  \frac{\sum_{l=1}^n R_{jl} \{(1-D_{jl}) \me^{X_l^\T \beta} +D_{jl} \me^{X_i^\T \beta}\} (X_i-X_l) \sum_{k=1}^n R_{jk} D_{jk} (X_i-X_k)^\T }
{(\sum_{l=1}^n R_{jl} \me^{X_l^\T \beta} )^2} . \label{eq:sigma-robins}
\end{align}
Although not apparent from the above definition, $\hat\sigma_j(\beta)$ can be equivalently expressed as a symmetric, nonnegative-definite matrix.
Moreover, $\tilde\sigma_j(\beta^*)$ can be shown to be conditionally unbiased for $\sigma_j(\beta^*)$, i.e.,
$ E \{ \hat \sigma_j(\beta^*)  | R_{j,1:n}, X_{1:n} \}= \sigma_j(\beta^*)$. See the Supplement for details.
However, in contrast with Proposition~\ref{pro:bet-model-based}(ii),
the sandwich variance $\hat \Sigma_{\text{b3}}$ does not automatically reduce to $\hat B^{-1} (\hat\beta)$,
the usual variance estimator for the maximum partial likelihood estimator, in the special case of no tied events.
A possible explanation is that $\hat\sigma_j (\beta)$ involves only two-way products of the event indicators $D_{ji}$,
whereas $\tilde\sigma_j(\beta)$ involves three-way products of the event indicators.

\vspace{.1in}
\noindent\textbf{Conditional inference given numbers of events}.
For odds ratio model (\ref{eq:odds-model2}), i.e., Cox's (1972) discrete-time propositional hazards model, a common approach for eliminating the nuisance parameters
$(\beta_{01}, \ldots, \beta_{0J})$ is to perform likelihood inference successively conditionally on the numbers of events $(T_1, \ldots, T_J)$, in addition to the risk-set indicators and covariates,
where $T_j = \sum_{i=1}^n R_{ji} D_{ji}$.
This approach is theoretically desirable (e.g., Lindsay 1980, 1983), but numerical implementation is intractable with a relatively large number of tied events.
Remarkably, we show that, given both the numbers of events and the risk-set indicators and covariates,
not only the individual terms, $\hat\tau_j(\beta)$, in the weighted Mantel--Haenszel estimating function are conditionally unbiased, but also
the variance estimators $\hat\sigma_j(\beta)$ evaluated at $\beta^*$ are conditionally unbiased.

\begin{pro} \label{pro:bet-model-based2}
Suppose that model (\ref{eq:odds-model2}) is correctly specified. For $j=1,\ldots, J$, each individual term $\hat\tau_j(\beta)$ is conditionally unbiased given $T_j$:
\begin{align}
E \{ \hat\tau_j(\beta^*) | T_j, R_{j,1:n}, X_{1:n} \} = 0, \label{eq:bet-unbiased-mean2}
\end{align}
where $T_j =\sum_{i=1}^n R_{ji} D_{ji}$, $R_{j,1:n} = (R_{j1},\ldots, R_{jn})$, and $X_{1:n}=(X_1, \ldots, X_n)$.
Moreover, $\hat \sigma_j(\beta^*) $ is conditionally unbiased for the conditional variance of $n^{1/2} \hat\tau_j(\beta^*)$:
\begin{align}
E \{ \hat \sigma_j(\beta^*)  | T_j, R_{j,1:n}, X_{1:n} \} = n\,\var\{ \hat\tau_j(\beta^*) | T_j, R_{j,1:n}, X_{1:n} \}. \label{eq:bet-unbiased-var2}
\end{align}
\end{pro}

There are two types of conditional unbiasedness,
depending on whether the risk-set indicators and covariates are conditioned on or
the number of events is further conditioned on.
See (\ref{eq:bet-unbiased-mean}) versus (\ref{eq:bet-unbiased-mean2}) for point estimation and
(\ref{eq:bet-unbiased-var}) and (\ref{eq:bet-unbiased-var2}) for variance estimation.
Based on Proposition~\ref{pro:bet-model-based2}, we expect that under suitable regularity conditions,
the point estimator $\hat\beta$ is consistent for $\beta^*$, and $n^{1/2} (\hat\beta-\beta^*)$ is asymptotically normal with mean 0 and
a variance matrix consistently estimated by the sandwich variance estimator $\hat\Sigma_{\text{b2}}$,
while conditioning on the number of events $(T_1,\ldots,T_J)$.
Large sample theory along this direction can be studied in future work.

Conditional unbiasedness given numbers of events, similar to (\ref{eq:bet-unbiased-mean2}),
is known to be satisfied by the Mantel--Haenszel estimating function for a common odds ratio in $2\times 2$ tables (Breslow 1981).
In that setting, conditional unbiasedness similar to (\ref{eq:bet-unbiased-var2}) is also established for the variance estimator in Robins et al.~(1986).
In fact, similarly to $\hat \sigma_j(\beta)$, the variance estimator $\tilde \sigma_j(\beta)$ in (\ref{eq:sigma-robins}) as an extension of Robins et al.~(1986) can also be shown to be conditionally unbiased,
that is, $E \{ \tilde \sigma_j(\beta^*)  | T_j, R_{j,1:n}, X_{1:n} \} = n\,\var\{ \hat\tau_j(\beta^*) | T_j, R_{j,1:n}, X_{1:n} \}$.
Nevertheless, the variance estimator $\hat \sigma_j(\beta)$ enjoys an exact reduction in the case of no tied events:
if $T_j=1$, then $ \hat\sigma_j(\beta^*) = n\,\var\{ \hat\tau_j(\beta^*) | T_j=1, R_{j,1:n}, X_{1:n} \}$, not just in expectation,
by Proposition~\ref{pro:bet-model-based}(ii) and the fact that the sample Hessian $\hat B(\beta^*)$ is equal to $ n\,\var\{ \hat\tau_j(\beta^*) | T_j=1, R_{j,1:n}, X_{1:n} \}$.

\vspace{.1in}
\noindent\textbf{Estimation of survival probabilities}.
Similarly as in Section~\ref{sec:prob-model}, we discuss estimation of survival probabilities for individuals with fixed covariates $x_0$. For simplicity, assume that $x_0=0$ in model (\ref{eq:odds-model}).
Then the hazard probability $p_j(x_0)$ is identified as $\expit(\beta_{0j})$, and can be estimated from (\ref{eq:bet0-est}) as
\begin{align*}
\hat q_j(x_0) = \expit(\hat\beta_{0j})= \frac{\sum_{i=1}^n R_{ji} D_{ji}} {\sum_{i=1}^n R_{ji} D_{ji} + R_{ji}(1-D_{ji}) \me^{ X_i^\T \hat\beta} }
\end{align*}
The $k$th survival probability, $P_k(x_0) = P( T > t_k |X=x_0)$, can be estimated as
\begin{align}
\hat Q_k( x_0) = \prod_{j=1}^k \{ 1- \hat q_j(x_0) \}  , \quad k=1,\ldots, J. \label{eq:bet-surv}
\end{align}
The estimators $\hat q_j(x_0)$ and $\hat Q_k(x_0)$ for $p_j(x_0)$ and $P_k(x_0)$
are automatically restricted to between 0 and 1, in contrast with $\hat p_j(x_0)$ and $\hat P_k(x_0)$ in Section~\ref{sec:prob-model}.
The cumulative hazard probability, $\sum_{j=1}^k p_j(x_0)$, can be estimated as $\sum_{j=1}^k \hat q_k(x_0)$.

The standard errors for $\hat Q_k(x_0)$ can be obtained using Taylor expansions (or the delta method) and either model-robust or model-based variance estimator for $\hat\beta$.
See the Supplement for detailed derivation and formulas.

\section{Comparison and extension} \label{sec:discussion}

\textbf{Pooled logistic regression}.\; For odds ratio model (\ref{eq:odds-model2}), i.e., Cox's (1972) discrete-time propositional hazard model,
conditional likelihood inference given numbers of events is usually considered statistically superior while exact solution can be numerically challenging.
For completeness, it is helpful to discuss another existing approach which directly uses maximum likelihood estimation over the main parameter $\beta$ and nuisance parameters $(\beta_{01},\ldots,\beta_{0J})$
in model (\ref{eq:odds-model2}) (e.g., Allison 1982).
The estimators, $\tilde\beta$ and $(\tilde\beta_{01}, \ldots, \tilde\beta_{0J})$, are defined jointly as a maximizer to the log likelihood function
\begin{align*}
\sum_{j=1}^J \sum_{i=1}^n R_{ji} \left\{ D_{ji} (\beta_{0j} + X^\T_i \beta) - \log \left( 1 + \me^{\beta_{0j} + X^\T_i \beta} \right) \right\} . 
\end{align*}
Equivalently, $\tilde\beta$ and $(\tilde\beta_{01}, \ldots, \tilde\beta_{0J})$ are determined jointly as a solution to
\begin{align}
& \sum_{i=1}^n R_{ji} \left\{ D_{ji} - \expit ( \beta_{0j} + X^\T_i \beta ) \right\} = 0, \quad j=1,\ldots, J,  \label{eq:Plogit-bet0} \\
& \sum_{j=1}^J \sum_{i=1}^n R_{ji} \left\{ D_{ji} - \expit (\beta_{0j} + X^\T_i \beta) \right\} X_i = 0. \label{eq:Plogit-bet}
\end{align}
This approach can be called pooled logistic regression, formally the same as fitting $J$ logistic regression models with a common coefficient vector $\beta$ across individual datasets.
On one hand, the estimating equations (\ref{eq:Plogit-bet0})--(\ref{eq:Plogit-bet}) are seemingly similar to estimating equation (\ref{eq:bet-est-a})--(\ref{eq:bet-est-b})
for weighted Mantel--Haenszel estimation in model (\ref{eq:odds-model2}), as well as (\ref{eq:gam-est-a})--(\ref{eq:gam-est-b}) for Breslow--Peto estimation in model (\ref{eq:prob-model2}).
On the other hand, there are fundamental differences between these methods which we explain as follows.

An easy difference is that closed-form solutions for $(\gamma_{01},\ldots,\gamma_{0J})$ from (\ref{eq:gam-est-a}) with fixed $\gamma$ or
for $(\beta_{01},\ldots,\beta_{0J})$ from (\ref{eq:bet-est-a}) with fixed $\beta$
can be derived, whereas such a closed-form solution is not available from equation (\ref{eq:Plogit-bet0}).
A deeper difference is that, to borrow the terminology of profile likelihood,
the profile estimating equation (\ref{eq:gam-est}) in $\gamma$ is conditionally unbiased according to (\ref{eq:gam-unbiased-mean}), and
the profile estimating equation (\ref{eq:bet-est}) in $\beta$ is conditionally unbiased according to (\ref{eq:bet-unbiased-mean}), both given the risk sets and covariates.
A profile estimating equation in $\beta$ can also be defined from equations (\ref{eq:Plogit-bet0})--(\ref{eq:Plogit-bet}), in spite of no closed-form solution for $(\beta_{01},\ldots,\beta_{0J})$.
But this estimating equation in $\beta$ does not satisfy conditional unbiasedness in a similar manner as (\ref{eq:gam-unbiased-mean}) or (\ref{eq:bet-unbiased-mean}).
Finally, the profile estimating equation (\ref{eq:bet-est}) in $\beta$ is also conditionally unbiased according to (\ref{eq:bet-unbiased-mean2}), given
the numbers of events in addition to the risk sets and covariates. This unbiasedness is shared by the conditional score equation in the approach of conditional likelihood inference.
For these reasons, weighted Mantel--Haeszel estimation is expected to achieve superior finite-sample performance, similarly as conditional likelihood estimation,
over pooled logistic regression, in particular with a large number of time points $J$.

The preceding discussion also explains that pooled logistic regression can be problematic in fitting model (\ref{eq:odds-model2})
with finely discretized data in finite samples, which is in agreement with the
understanding that maximum likelihood estimation with a large number of nuisance parameters may not generally be desirable.

\vspace{.1in}
\textbf{Time-varying coefficient and time-dependent covariates}.\;
Our theory and methods are so far developed in the context of models (\ref{eq:prob-model}) and (\ref{eq:odds-model}),
with time-independent regression coefficients and time-independent covariates.
Nevertheless, the development can be readily extended to handle time-varying coefficients and time-dependent covariates,
similarly as in Cox's continuous-time proportional hazards models.
First, consider an extension of models (\ref{eq:prob-model}) and (\ref{eq:odds-model}),
where $p_j(x)$ is redefined as
\begin{align*}
p_j(x) = P(Y=t_j , \delta= 1 | Y \ge t_j , X(t_j) = x ),
\end{align*}
where $X(t_j)$ is the covariate vector at time $t_j$. Then estimating equations (\ref{eq:gam-est}) for $\hat\gamma$
and (\ref{eq:bet-est}) for $\hat\beta$ can be extended by replacing $X_i$ with $X_i(t_j)$ within the $j$th risk set.
Similar modification can be applied to the model-based and model-robust variance estimators.
Next, time-varying coefficients can be accommodated by a reformulation using time-dependent covariates.
For example, consider model (\ref{eq:prob-model}) extended with a time-varying coefficient for a scalar covariate $x^{(1)}$:
\begin{align}
p_j (x) = p_j (x_0) \exp \left\{  x^{(1)} b(t_j; \gamma^*_{11}, \gamma^*_{12} ) + x^{(2)\T} \gamma^*_2 \right\}, \label{eq:prob-model3}
\end{align}
where $x_0=0$, $x=(x^{(1)}, x^{(2)\T})^\T$, $b(t; \gamma_{11}, \gamma_{12} )$ is a function of time, defined as $\gamma_1 + u^\T(t) \gamma_{12}$ using
a basic vector $u(t)$, and $\gamma^* = (\gamma^*_{11}, \gamma^{*\T}_{12}, \gamma_2^{*\T})^\T$ are unknown coefficients.
Model (\ref{eq:prob-model3}) can be put in the form of (\ref{eq:prob-model}), where $x$ is replaced by the time-dependent covariate vector $(x^{(1)}, x^{(1)} u^\T (t_j), x^{(2)\T})^\T$ at time $t_j$
associated with the coefficient vector $\gamma^*$.

\section{Numerical studies}

\subsection{Analysis of veteran's lung cancer data} \label{sec:VA}

We compare different methods in analysis of the data on a Veteran's Administration lung cancer trial used in Kalbflwisch \& Prentice (1980).
The trial included 137 male patients with advanced lung cancer. The outcome of interest is time to death in days, and there are six covariates
measured at randomization: treatment (test or standard), age in years, Karnofsky score (ranged 10 to 99), time in months from diagnosis to the start of treatment,
cell type (a nominal factor of 4 levels), and prior therapy (yes or no).
The corresponding regression terms are denoted as \texttt{treat}, \texttt{age}, \texttt{Karn}, \texttt{diagt}, \texttt{cell2, cell3, cell4}
(for the contrasts between levels 2--4 versus 1), and \texttt{prior}.

Kaplan--Meier survival curves suggest non-proportional hazards over time in the two treatment groups, while ignoring other covariates (Tan 2019, Supplement).
Hence we fit hazard probability and odds models by allowing time-varying coefficients with the treatment variable.
As discussed in Section \ref{sec:discussion}, such models can be stated using time-dependent covariates (or regression terms), defined as functions of the time and treatment variables, \texttt{time} and \texttt{treat}.
For simplicity, we include two time-dependent regression terms, \texttt{treat2} and \texttt{treat3}, defined as
$\texttt{treat}*1\{\texttt{time}>100\}$ and $\texttt{treat} *1\{\texttt{time}>200\}$.
The coefficients for these two terms represent changes  after day 100 or 200 in the association of the test treatment with hazard probabilities or odds.

To study discrete-time inference, we also apply various methods to further discretized data, obtained by grouping the original times in intervals of 20 days.
For concreteness, the censored-late option is used as mentioned in Section~\ref{sec:data}. An uncensored time in $(t_{j-1},t_j]$ is labeled $t_j$, whereas
a censored time in $[t_{j-1},t_j)$ is labeled $t_j$. The censoring indicator is kept unchanged. See Tan (2019, Supplement) for more details.

\begin{table}
\caption{Analysis of veteran's lung cancer data (original)} \label{tab:VA-contT}
\footnotesize
\begin{center} \renewcommand\arraystretch{.9}
\begin{tabular*}{\textwidth}{c rrrrr c rrrrr} \hline
    &   BP  & Efron  &   CML  &  wMH    & Plogit   &  &  BP  & Efron  &   CML  &  wMH    & Plogit \\
    & \multicolumn{5}{c}{Point estimate}  &  & \multicolumn{5}{c}{Point estimate} \\ \cline{2-6} \cline{8-12}

treat  & $.379$ & $.383$ & $.384$ & $.383$ & $.392$  &       diagt  & $-.064$ & $-.047$ & $-.060$ & $-.038$ & $-.080$ \\
treat2 & $-.493$ & $-.494$ & $-.498$ & $-.494$ & $-.511$  &  cell2  & $.830$ & $.835$ & $.836$ & $.830$ & $.865$ \\
treat3 & $.472$ & $.476$ & $.483$ & $.475$ & $.437$  &       cell3  & $1.152$ & $1.161$ & $1.167$ & $1.167$ & $1.196$ \\
age    & $-.813$ & $-.829$ & $-.813$ & $-.838$ & $-.804$  &  cell4  & $.372$ & $.374$ & $.376$ & $.376$ & $.385$ \\
Karn   & $-.320$ & $-.322$ & $-.324$ & $-.323$ & $-.334$ &   prior  & $.083$ & $.083$ & $.084$ & $.087$ & $.082$ \\[2ex] \hline

    & oldBP   & BP  & Efron  &   CML  &  wMH    & Plogit   & BP  & Efron  &  CML  & wMH    & Plogit   \\
    & \multicolumn{6}{c}{Model-based SE}  & \multicolumn{4}{c}{Model-robust SE}  \\  \cline{2-7} \cline{8-12}

treat  & $.245$ & $.243$ & $.245$ & $.246$ & $.247$ & $.248$ & $.221$ & $.223$ & --- & $.224$ & $.227$ \\
treat2 & $.516$ & $.515$ & $.516$ & $.518$ & $.515$ & $.524$ & $.481$ & $.484$ & --- & $.482$ & $.496$ \\
treat3 & $.645$ & $.645$ & $.646$ & $.647$ & $.644$ & $.670$ & $.622$ & $.624$ & --- & $.622$ & $.662$ \\
age    & $.931$ & $.927$ & $.930$ & $.937$ & $.930$ & $.954$ & $1.029$ & $1.036$ & --- & $1.035$ & $1.082$ \\
Karn   & $.056$ & $.056$ & $.056$ & $.057$ & $.056$ & $.058$ & $.053$ & $.054$ & --- & $.054$ & $.057$ \\
diagt  & $.918$ & $.897$ & $.919$ & $.930$ & $.947$ & $.945$ & $.790$ & $.790$ & --- & $.800$ & $.833$ \\
cell2  & $.283$ & $.282$ & $.283$ & $.284$ & $.284$ & $.288$ & $.306$ & $.309$ & --- & $.310$ & $.321$ \\
cell3  & $.313$ & $.311$ & $.313$ & $.315$ & $.315$ & $.319$ & $.273$ & $.275$ & --- & $.277$ & $.284$ \\
cell4  & $.292$ & $.291$ & $.292$ & $.293$ & $.292$ & $.297$ & $.247$ & $.248$ & --- & $.248$ & $.258$ \\
prior  & $.232$ & $.231$ & $.232$ & $.233$ & $.234$ & $.238$ & $.217$ & $.219$ & --- & $.220$ & $.226$ \\ \hline

\end{tabular*} \\[1ex]
\parbox{\textwidth}{\scriptsize Note:
BP, wMH, or Plogit denotes Breslow--Peto estimator $\hat\gamma$, weighted Mantel--Haenszel estimator $\hat\beta$, or pooled logistic estimator $\tilde\beta$,
implemented by the R package \texttt{dSurvival} (Tan 2020b).
oldBP, Efron, or CML denotes results from Cox's regression \texttt{coxph} with \texttt{ties="breslow"}, \texttt{"efron"}, or \texttt{"exact"} in the R package \texttt{survival} (Therneau 2015).
oldBP and BP are identical to each other in point estimates and model-robust SEs.
The point estimates and SEs for \texttt{age}, \texttt{Karn}, and \texttt{diagt} are reported after multiplied by $10^2$, $10$, and $10^2$ respectively.}
\end{center} \vspace{-.2in}
\end{table}

\begin{table}
\caption{Analysis of veteran's lung cancer data (discretized)} \label{tab:VA-contF}
\footnotesize
\begin{center} \renewcommand\arraystretch{.9}
\begin{tabular*}{\textwidth}{c rrrrr c rrrrr} \hline
    &   BP  & Efron  &   CML  &  wMH    & Plogit   &  &  BP  & Efron  &   CML  &  wMH    & Plogit \\
    & \multicolumn{5}{c}{Point estimate}  &  & \multicolumn{5}{c}{Point estimate} \\ \cline{2-6} \cline{8-12}

treat  & $.307$ & $.346$ & $.415$ & $.420$ & $.422$       & diagt & $-.007$ & $-.129$ & $.048$ & $.040$ & $.034$ \\
treat2 & $-.476$ & $-.463$ & $-.567$ & $-.484$ & $-.581$ & cell2 & $.778$ & $.859$ & $.926$ & $.916$ & $.955$ \\
treat3 & $.419$ & $.437$ & $.546$ & $.406$ & $.507$      & cell3 & $1.047$ & $1.159$ & $1.365$ & $1.382$ & $1.393$ \\
age    & $-.459$ & $-.744$ & $-.362$ & $-.754$ & $-.343$  & cell4 & $.366$ & $.408$ & $.464$ & $.517$ & $.473$ \\
Karn   & $-.267$ & $-.310$ & $-.358$ & $-.337$ & $-.368$  & prior & $.053$ & $.103$ & $.061$ & $.079$ & $.056$ \\[2ex] \hline

    & oldBP   & BP  & Efron  &   CML  &  wMH    & Plogit   & BP  & Efron  &  CML  & wMH    & Plogit   \\
    & \multicolumn{6}{c}{Model-based SE}  & \multicolumn{4}{c}{Model-robust SE}  \\  \cline{2-7} \cline{8-12}

treat & $.241$ & $.204$ & $.244$ & $.273$ & $.305$ & $.275$ & $.191$ & $.219$ & ---  & $.264$ & $.244$ \\
treat2 & $.514$ & $.473$ & $.516$ & $.554$ & $.570$ & $.561$ & $.452$ & $.488$ & --- & $.528$ & $.533$ \\
treat3 & $.645$ & $.611$ & $.645$ & $.684$ & $.694$ & $.707$ & $.600$ & $.628$ & --- & $.669$ & $.709$ \\
age   & $.920$ & $.794$ & $.923$ & $1.055$ & $1.087$ & $1.072$ & $.924$ & $1.039$ & --- & $1.216$ & $1.250$ \\
Karn  & $.054$ & $.047$ & $.055$ & $.066$ & $.063$ & $.067$ & $.046$ & $.053$ & --- & $.060$ & $.068$ \\
diagt & $.925$ & $.746$ & $.931$ & $1.192$ & $1.173$ & $1.205$ & $.704$ & $.786$ & --- & $.925$ & $1.139$ \\
cell2 & $.279$ & $.250$ & $.281$ & $.306$ & $.327$ & $.310$ & $.270$ & $.300$ & --- & $.348$ & $.342$ \\
cell3 & $.309$ & $.269$ & $.312$ & $.351$ & $.375$ & $.355$ & $.236$ & $.265$ & --- & $.302$ & $.291$ \\
cell4 & $.291$ & $.270$ & $.292$ & $.311$ & $.324$ & $.315$ & $.224$ & $.243$ & --- & $.261$ & $.264$ \\
prior & $.232$ & $.205$ & $.234$ & $.257$ & $.272$ & $.262$ & $.196$ & $.223$ & --- & $.247$ & $.248$ \\ \hline

\end{tabular*} 
\end{center} \vspace{-.2in}
\end{table}

\begin{figure}
\begin{tabular}{c}
\includegraphics[width=6.2in, height=2.5in]{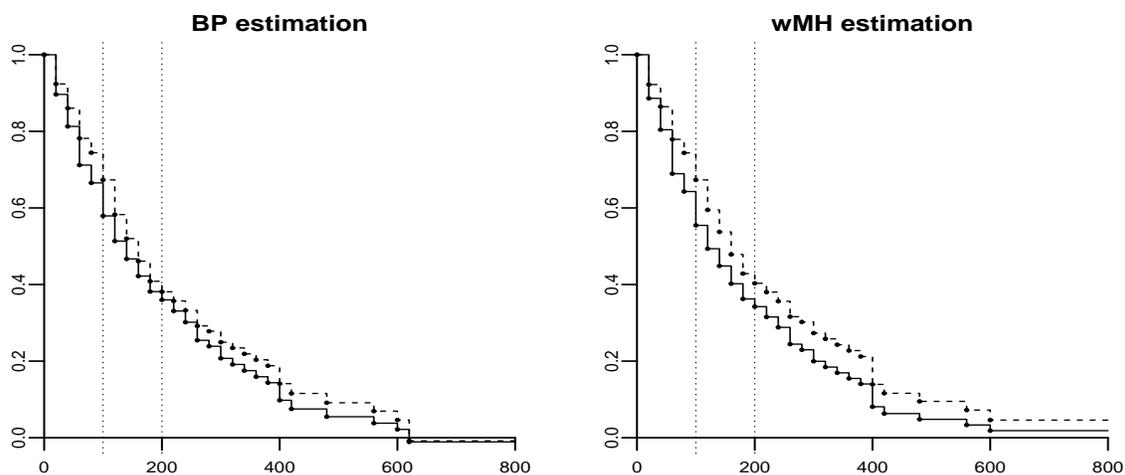}
\end{tabular}
\caption{\small Survival probabilities for individuals in the test (solid) or standard (dashed) treatment group, with covariates
\texttt{age}$=60$, \texttt{Karn}$=60$, \texttt{diagt}$=9$, \texttt{cell}$=1$, and \texttt{prior}$=0$. Two vertical lines are placed at days 100 and 200.}
\label{fig:VA-disc} \vspace{.1in}
\end{figure}

Tables~\ref{tab:VA-contT}--\ref{tab:VA-contF} present the results on the original and discretized data.
For the original data with a small number of tied deaths, the estimates of BP, Efron, CML, and wMH are similar to each other in various degrees,
although the BP point estimates associated with probability ratios are consistently closer to 0 than those of CML and wMH associated with odds ratios,
except for the coefficient of \texttt{diagt} which is the least accurately estimated
as measured by the $t$-statistic. The Plogit point estimates show noticeable differences (or biases) from those of CML and wMH.

For the discretized data with more tied deaths, the BP point estimates
are more substantially closer to 0 than those of CML and wMH, which remain similar to each other at least for coefficients with relatively large $t$-statistics.
This difference can be properly explained by the fact that BP estimates are associated with odds ratios,
whereas the CML and wMH estimates are associated with probability ratios.
In addition, in a more pronounced manner than in Table~\ref{tab:VA-contT},
the commonly reported variance estimates in the column oldBP are inflated compared with the proposed variance estimates in the column BP, as expected by Corollary~\ref{cor:gam-bres-var}.
For example, for the coefficient of \texttt{treat}, the BP point estimate is smaller than CML by $1- .307/.415 = 26.0\%$,
and the oldBP variance estimate is larger than the proposed BP variance estimate by $(.241/ .204)^2 -1= 39.6\%$.
The Efron estimates tend to fall between BP and CML estimates. The Plogit point estimates still show
various differences from those of CML and wMH.

For illustration, Figure~\ref{fig:VA-disc} shows the estimated survival probabilities using the BP and wMH methods with the discretized data, for individuals in the
test or standard treatment group and with certain fixed covariate values. As allowed by the specified models, the test treatment compared with the standard treatment
is associated with increasingly lower survival probabilities over time before day 100 or after day 200, while the trend is reversed between day 100 and 200.
The BP estimate of the last survival probability is negative, a possibility mentioned in Section~\ref{sec:prob-model}.
This also reflects the fact that such estimates in the right tail are usually inaccurate.

\begin{table}
\caption{Comparison from simulated data (finely discretized, proportional hazards)} \label{tab:sim-contT-phT}
\footnotesize
\begin{center} \renewcommand\arraystretch{.9}
\begin{tabular*}{.9\textwidth}{ll rrrrr rrrrr} \hline
    && BP  & Efron  &   CML  &  wMH    & Plogit   & BP  & Efron  &   CML  &  wMH    & Plogit   \\
    &&   \multicolumn{5}{c}{Point mean}  &   \multicolumn{5}{c}{Point SD} \\ \cline{2-12}

Tr && $-.408$ & $-.411$ & $-.413$ & $-.413$ & $-.427$ & $.259$ & $.260$ & $.261$ & $.262$ & $.271$ \\
X1 && $.626$ & $.629$ & $.633$ & $.633$ & $.655$ & $.164$ & $.166$ & $.167$ & $.167$ & $.174$ \\
X2 && $-.412$ & $-.414$ & $-.416$ & $-.416$ & $-.430$ & $.177$ & $.178$ & $.179$ & $.179$ & $.186$ \\
X3 && $.305$ & $.307$ & $.308$ & $.309$ & $.319$ & $.173$ & $.174$ & $.175$ & $.175$ & $.181$ \\
X4 && $.107$ & $.108$ & $.109$ & $.109$ & $.112$ & $.152$ & $.153$ & $.154$ & $.154$ & $.160$ \\ [2ex] \hline

    && oldBP   & BP  & Efron  &   CML  &  wMH    & Plogit   & BP  & Efron  &  wMH    & Plogit   \\
    && \multicolumn{6}{c}{Model-based SE}  & \multicolumn{4}{c}{Model-robust SE}  \\       \cline{2-12}

Tr && $.250$ & $.249$ & $.250$ & $.252$ & $.252$ & $.257$ & $.243$ & $.245$ & $.246$ & $.256$ \\
X1 && $.160$ & $.159$ & $.160$ & $.161$ & $.162$ & $.165$ & $.155$ & $.156$ & $.157$ & $.164$ \\
X2 && $.168$ & $.167$ & $.168$ & $.169$ & $.169$ & $.173$ & $.162$ & $.163$ & $.164$ & $.170$ \\
X3 && $.166$ & $.165$ & $.166$ & $.167$ & $.167$ & $.170$ & $.160$ & $.161$ & $.162$ & $.169$ \\
X4 && $.145$ & $.144$ & $.145$ & $.146$ & $.146$ & $.149$ & $.140$ & $.141$ & $.142$ & $.147$ \\ \hline
\end{tabular*} \\[1ex]
\parbox{\textwidth}{\scriptsize Note: See the footnote for Table~\ref{tab:VA-contT}.
Point mean and SD are the Monte Carlo mean and standard deviation of the point estimates, and model-based and model-robust SEs
are the square roots of the Monte Carlo mean of the model-based and model-robust variance estimates. }
\end{center} \vspace{-.2in}
\end{table}

\begin{table}
\caption{Comparison from simulated data (coarsely discretized, proportional hazards)} \label{tab:sim-contF-phT}
\footnotesize
\begin{center} \renewcommand\arraystretch{.9}
\begin{tabular*}{.9\textwidth}{ll rrrrr rrrrr} \hline
    && BP  & Efron  &   CML  &  wMH    & Plogit   & BP  & Efron  &   CML  &  wMH    & Plogit   \\
    &&   \multicolumn{5}{c}{Point mean}  &   \multicolumn{5}{c}{Point SD} \\ \cline{2-12}

Tr && $-.353$ & $-.389$ & $-.432$ & $-.437$ & $-.444$ & $.234$ & $.258$ & $.286$ & $.293$ & $.295$ \\
X1 && $.540$ & $.596$ & $.663$ & $.674$ & $.683$ & $.145$ & $.162$ & $.184$ & $.192$ & $.191$ \\
X2 && $-.356$ & $-.393$ & $-.436$ & $-.444$ & $-.450$ & $.158$ & $.176$ & $.197$ & $.204$ & $.204$ \\
X3 && $.263$ & $.290$ & $.322$ & $.328$ & $.332$ & $.156$ & $.173$ & $.191$ & $.199$ & $.198$ \\
X4 && $.093$ & $.103$ & $.114$ & $.116$ & $.118$ & $.137$ & $.152$ & $.168$ & $.174$ & $.174$ \\ [2ex] \hline

    && oldBP   & BP  & Efron  &   CML  &  wMH    & Plogit   & BP  & Efron  &  wMH    & Plogit   \\
    && \multicolumn{6}{c}{Model-based SE}  & \multicolumn{4}{c}{Model-robust SE}  \\       \cline{2-12}

Tr && $.248$ & $.225$ & $.249$ & $.274$ & $.283$ & $.280$ & $.221$ & $.242$ & $.273$ & $.279$ \\
X1 && $.156$ & $.139$ & $.158$ & $.178$ & $.190$ & $.182$ & $.137$ & $.153$ & $.179$ & $.181$ \\
X2 && $.165$ & $.148$ & $.167$ & $.185$ & $.195$ & $.189$ & $.145$ & $.160$ & $.184$ & $.187$ \\
X3 && $.164$ & $.147$ & $.165$ & $.182$ & $.191$ & $.186$ & $.144$ & $.159$ & $.181$ & $.184$ \\
X4 && $.144$ & $.129$ & $.145$ & $.159$ & $.166$ & $.162$ & $.127$ & $.140$ & $.158$ & $.161$ \\ \hline
\end{tabular*} 
\end{center} 
\end{table}

\subsection{Simulation study} \label{sec:simulation}

To further compare different methods, we also conduct simulation studies. The first study, reported below, involves simulated data satisfying proportional hazards in continuous time,
whereas the second study, reported in the Supplement, involves simulated data where proportional hazards are violated even in continuous time.

For each simulation, a sample of size $n=100$ is generated as follows, mimicking a randomized trial.
The treatment variable \texttt{Tr} is generated as 1 (test) or 2 (standard) with probabilities $.5$ each, and
four covariates  \texttt{X1}--\texttt{X4} are generated, independently of \texttt{Tr}, as multivariate normal with
means 0 and covariances $2^{-|j-k|}$ between $j$th and $k$th covariates for $1\le j, k\le 4$.
The event time $\tilde T$ is generated as Exponential with scale parameter $\exp(- X^\T \beta^*)$,
where $X$ consists of \texttt{Tr} and \texttt{X1}--\texttt{X4} and $\beta^*= (-.4, .6, -.4, .3, .1)^\T$.
The censoring variable $\tilde C$ is generated as Uniform between 0 and $4 \exp(- X^\T \beta^*)$.
To study discrete-time inference, two sets of observed data $(Y, \delta$) are obtained, where $\delta = 1 \{\tilde T \le \tilde C\}$
and $Y$ is defined by discretizing $\tilde Y = \min(\tilde T, \tilde C)$ in intervals of length $.01$ or $.2$,
using the censor-late option. Both probability model (\ref{eq:prob-model}) and odds model (\ref{eq:odds-model})
are fit with the regression terms \texttt{Tr} and \texttt{X1}--\texttt{X4}.
These models are misspecified, to a much less extent for the finely discretized data than for the coarsely discretized data.
Nevertheless, inference can be performed by treating these models as approximations.

Table~\ref{tab:sim-contT-phT}--\ref{tab:sim-contF-phT} present the results from 2000 repeated simulations.
There are similar patterns in these results as  in Tables~\ref{tab:VA-contT}--\ref{tab:VA-contF}.
The point estimates are close to the true values in $\beta^*$ for finely discretized data, with Plogit the most biased.
For coarsely discretized data, the BP estimates are attenuated from $\beta^*$ toward 0, whereas the CML and wMH are amplified away from 0,
by the nature of how these estimators are associated with probability or odds ratios.
In addition, all the model-based and model-robust variance estimates appear to
reasonably match the Monte Carlo variances, regardless of theoretical consistency.
Such agreement between model-based and model-robust variance estimation may not generally hold.
Nevertheless, various degrees of under-estimation can be found from these variance estimates,
except the commonly reported model-based variance estimates in the column oldBP, which are upward biased,
for example, by $(.156/.145)^2-1=15.7\%$ for the coefficient of \texttt{X1} with coarsely discretized data.

\section{Conclusion}

For discrete-time survival analysis, we develop new methods and theory using numerically simple and conditionally unbiased estimating functions, along with
model-based and model-robust variance estimation, in hazard probability and odds models.
The latter is known as Cox's discrete-time proportional hazards model.
Due to conditional unbiasedness, our methods are expected to perform
satisfactorily in a broad range of settings, with small or large numbers of tied events corresponding to a large or small number of time intervals.
In fact, the Breslow--Peto and the weighted Mantel--Haenszel estimators and the associated model-based variances estimators
reduce to the partial likelihood estimator and the associated variance estimator in the extreme case of only one event per risk set
as would be observed in the continuous-time setting.
In this sense, our work provides unified methods for both discrete- and continuous-time survival analysis.
Similar ideas can be pursued to address other related problems.

\section{Appendix: Variance estimation for survival probabilities}

The standard error (SE) for the estimated survival probability $\hat P_k(x_0)$ can be obtained using the delta method as $\hat P_k(x_0) \,\text{SE} \{ \log \hat P_k(x_0)\}$.
Therefore, it suffices to determine $\text{SE} \{ \log \hat P_k(x_0)\}$.
For simplicity, assume that all $J$ risk sets are sufficiently large.

\subsection{Hazard probability model}

First, we derive model-robust variance estimation for $ \log \hat P_k(x_0)$, where $\hat P_k( x_0) = \prod_{j=1}^k ( 1- \me^{\hat\gamma_{0j}} )$ from (\ref{eq:gam-surv}).
Consider the Taylor expansion
\begin{align*}
& \log \hat P_k(x_0) - \log \bar P_k(x_0) = \sum_{j=1}^k \left\{ \log \left( 1- \me^{\hat\gamma_{0j}} \right) - \log \left( 1- \me^{\bar\gamma_{0j}} \right) \right\} \\
& =  \sum_{j=1}^k \frac{-1}{1-\me^{\bar\gamma_{0j}}} \left( \me^{\hat\gamma_{0j}} - \me^{\bar\gamma_{0j}} \right) + o_p(n^{-1/2}),
\end{align*}
where $\bar P_k(x_0) = \prod_{j=1}^k ( 1- \me^{\bar\gamma_{0j}} )$ and
$ \me^{\bar\gamma_{0j}} = E (\tilde R_j \tilde D_j) /E( \tilde R_j \me^{\tilde X^\T \bar\gamma})$, which is the probability limit of $\me^{\hat\gamma_{0j}}$. Then we use the decomposition
\begin{align*}
\me^{\hat\gamma_{0j}} - \me^{\bar\gamma_{0j}}  =
\left\{ \frac{\sum_{i=1}^n R_{ji} D_{ji}} {\sum_{i=1}^n R_{ji} \me^{ X_i^\T \bar\gamma}} - \me^{\bar\gamma_{0j}} \right\} +
\left\{ \frac{\sum_{i=1}^n R_{ji} D_{ji}} {\sum_{i=1}^n R_{ji} \me^{ X_i^\T \hat\gamma}} -
\frac{\sum_{i=1}^n R_{ji} D_{ji}} {\sum_{i=1}^n R_{ji} \me^{ X_i^\T \bar\gamma}}  \right\}.
\end{align*}
The first term can be approximated as
\begin{align}
& \frac{\sum_{i=1}^n R_{ji} D_{ji}} {\sum_{i=1}^n R_{ji} \me^{ X_i^\T \bar\gamma}} - \me^{\bar\gamma_{0j}} =
  \frac{ \sum_{i=1}^n R_{ji} ( D_{ji} - \me^{\bar\gamma_{0j} + X^\T_i \bar\gamma} ) } {\sum_{i=1}^n R_{ji} \me^{ X_i^\T \bar\gamma}} \nonumber  \\
& = \frac{ n^{-1} \sum_{i=1}^n R_{ji} ( D_{ji} - \me^{\bar\gamma_{0j} + X^\T_i \bar\gamma} ) } {E (\tilde R_j \me^{\tilde X^\T \bar\gamma} ) } + o_p(n^{-1/2}) . \label{eq:gam-surv-first-term}
\end{align}
The second term can be approximated as
\begin{align*}
& \frac{\sum_{i=1}^n R_{ji} D_{ji}} {\sum_{i=1}^n R_{ji} \me^{ X_i^\T \hat\gamma}} - \frac{\sum_{i=1}^n R_{ji} D_{ji}} {\sum_{i=1}^n R_{ji} \me^{ X_i^\T \bar\gamma}}
=  \frac{\sum_{i=1}^n R_{ji} D_{ji}} {\sum_{i=1}^n R_{ji} \me^{ X_i^\T \hat\gamma} } \frac{\sum_{i=1}^n R_{ji}  (\me^{ X_i^\T \bar\gamma} - \me^{ X_i^\T \hat\gamma} )}
{\sum_{i=1}^n R_{ji} \me^{ X_i^\T \bar\gamma} }\\
& = \frac{- E(\tilde R_j \tilde D_j) E(\tilde R_j \me^{\tilde X^\T \bar\gamma} \tilde X^\T) } { E^2 (\tilde R_j \me^{ \tilde X^\T \bar\gamma}) } (\hat\gamma-\bar\gamma) + o_p(n^{-1/2}) \\
& =  \frac{- E(\tilde R_j \tilde D_j) E(\tilde R_j \me^{\tilde X^\T \bar\gamma} \tilde X^\T) } { E^2 (\tilde R_j \me^{ \tilde X^\T \bar\gamma}) } B^{-1}(\bar\gamma) \frac{1}{n} \sum_{i=1}^n h_\bullet (Y_i, \delta_i, X_i; \bar\gamma)+ o_p(n^{-1/2}),
\end{align*}
where $h_\bullet (Y,\delta,X;\gamma) =\sum_{j=1}^J h_j(Y,\delta,X;\gamma)$.
Combining the preceding four displays yields the asymptotic expansion
\begin{align}
\log \hat P_k(x_0) - \log \bar P_k(x_0) = \frac{1}{n} \sum_{i=1}^n \varphi_k (Y_i,\delta_i,X_i; \bar\gamma) + o_p(n^{-1/2}), \label{eq:gam-surv-expan}
\end{align}
where, with $\me^{\gamma_{0j}}$ set to $E (\tilde R_j \tilde D_j) /E( \tilde R_j \me^{\tilde X^\T \gamma})$,
\begin{align*}
& \varphi_k(Y,\delta,X; \gamma) \\
& =\sum_{j=1}^k \frac{-1}{1-\me^{\gamma_{0j}}} \left\{ \frac{R_j ( D_j - \me^{\gamma_{0j} + X^\T \gamma} )}{E (\tilde R_j \me^{\tilde X^\T \gamma} )}  -
\frac{E(\tilde R_j \tilde D_j) E(\tilde R_j \me^{\tilde X^\T \gamma} \tilde X^\T) } {E^2 (\tilde R_j \me^{ \tilde X^\T \gamma}) } B^{-1}(\gamma)h_\bullet (Y, \delta, X; \gamma)\right\} .
\end{align*}
Then a model-robust variance estimator for $\log \hat P_k(x_0)$ is  $n^{-2} \sum_{i=1}^n  \hat\varphi_k^2 (Y_i,\delta_i,X_i; \hat\gamma)$, where
$\hat \varphi_k (Y,\delta,X; \gamma)$ is defined as $\varphi_k (Y,\delta, X;\gamma)$ with
$B(\gamma)$ replaced by $\hat B(\gamma)$,
$h_\bullet (Y, \delta, X; \gamma)$ replaced by $\hat h_\bullet (Y, \delta, X;\gamma) =\sum_{j=1}^J \hat h_j(Y,\delta,X;\gamma)$, and
$E (\tilde R_j \tilde D_j)$, $E(\tilde R_j \me^{ \tilde X^\T \gamma})$, and $E( \tilde R_j \me^{\tilde X^\T \gamma} \tilde X )$ replaced by the corresponding sample averages.

For model-based variance estimation, suppose that model (\ref{eq:prob-model2}) is correctly specified and hence $\bar\gamma=\gamma^*$.
We return to the asymptotic expansion (\ref{eq:gam-surv-expan}), and use the fact that
the individual terms $R_j ( D_j - \me^{\gamma^*_{0j} + X^\T \gamma^*})$ are uncorrelated not only with each other for $j=1,\ldots,J$
but also with $ h_\bullet(Y, \delta, X; \gamma^*)$. The asymptotic variance of $\log \hat P_k(x_0)$ can be simplified as
\begin{align}
\frac{1}{n} \left[ \sum_{j=1}^k  \frac{E \{R_j p_j(X) (1-p_j(X))\}}{ (1-\me^{\gamma^*_{0j}})^2 E^2 (\tilde R_j \me^{\tilde X^\T \gamma^*} ) }
+ U_k^\T V  U_k \right], \label{eq:gam-surv-var}
\end{align}
where  $p_j(X) = \me^{\gamma^*_{0j} + X^\T \gamma^* } $, $\me^{\gamma^*_{0j}}=
E( \tilde R_j\tilde D_j) /E(\tilde R_j \me^{ \tilde X^\T \gamma^*})$, and
\begin{align*}
U_k = \sum_{j=1}^k \frac{1}{1-\me^{\gamma^*_{0j}}}\frac{E(\tilde R_j \tilde D_j) E(\tilde R_j \me^{\tilde X^\T \gamma^*} \tilde X) } { E^2 (\tilde R_j \me^{ \tilde X^\T \gamma^*})} .
\end{align*}
A model-based variance estimator for $\log \hat P_k(x_0)$ is obtained from (\ref{eq:gam-surv-var}) with $\gamma^*$ replaced by $\hat\gamma$, $V$ replaced by
$\hat V_{\text{b}}$ or $\hat V_{\text{b2}}$, and
all expectations replaced by the corresponding sample averages, for example, the expectation $E \{R_j p_j(X) (1-p_j(X))\}$ replaced by
$n^{-1} \sum_{i=1}^n R_j \hat p_j(X_i;\hat\gamma) (1-\hat p_j(X_i; \hat\gamma))$,
where $\hat p_j(X;\gamma) =
\{(\sum_{l=1}^n R_{jl} D_{jl}) / (\sum_{l=1}^n R_{jl} \me^{ X_l^\T \gamma})\}  \me^{X^\T \gamma}$ as in the definition of $\hat A_{\text{b}}(\gamma)$.
For the R package \texttt{survival} when using the Breslow--Peto estimator, the model-based variance estimator for the cumulative hazard $\sum_{j=1}^k \me^{\hat\gamma_{0j}}$,
or for the estimator $\log \hat P^\dag_k(x_0)$
with $\hat P^\dag_k(x_0) = \me^{-  \sum_{j=1}^k \me^{\hat\gamma_{0j}}}$, is computed from
\begin{align}
& \frac{1}{n} \left[ \sum_{j=1}^k  \frac{E (\tilde R_j \tilde D_j )}{ E^2 (\tilde R_j \me^{\tilde X^\T \gamma^*} ) }  + U^{\dag\T}_k V  U^\dag_k \right]
=\frac{1}{n} \left[ \sum_{j=1}^k  \frac{E \{R_j p_j(X) \}}{ E^2 (\tilde R_j \me^{\tilde X^\T \gamma^*} )  }  + U^{\dag\T}_k V  U^\dag_k \right], \label{eq:gam-surv-var-alt}
\end{align}
with \vspace{-.2in}
\begin{align*}
U^\dag_k = \sum_{j=1}^k \frac{E(\tilde R_j \tilde D_j) E(\tilde R_j \me^{\tilde X^\T \gamma^*} \tilde X) } { E^2 (\tilde R_j \me^{ \tilde X^\T \gamma^*})} .
\end{align*}
by replacing $\gamma^*$ with $\hat\gamma$, $V$ with
$\hat B^{-1}(\hat\gamma)$, and
all expectations replaced with the corresponding sample averages (Therneau \& Grambsch 2000, Section 10.2.3).
Even after ignoring the factor $1-\me^{\gamma^*_{0j}}$, there are two important differences between the two variance estimators based on (\ref{eq:gam-surv-var}) and (\ref{eq:gam-surv-var-alt}):
the term $E \{R_j \hat p_j(X;\hat\gamma) (1-\hat p_j(X;\hat\gamma))\}$ versus $E \{R_j \hat p_j(X;\hat\gamma) \}$
and the variance estimator $\hat V_{\text{b}}$ or $\hat V_{\text{b2}}$ versus $\hat B^{-1}(\hat\gamma)$.

Incidentally, it appears by numerical evaluation that for the R package \texttt{survival} when using the Breslow--Peto estimator,
the model-robust variance estimator for $\log \hat P^\dag_k(x_0)$
with $\hat P^\dag_k(x_0) = \me^{-\sum_{j=1}^k \me^{\hat\gamma_{0j}}}$, is computed from (\ref{eq:gam-surv-var-alt}) similarly as
the model-based variance estimator, except that $V$ is replaced by $\hat V_{\text{r}}$ instead of $\hat B^{-1}(\hat\gamma)$.
This method is theoretically problematic, because if model (\ref{eq:odds-model2}) is misspecified, then
the asymptotic variance for $\sum_{j=1}^k \me^{\hat\gamma_{0j}}$ does not in general admit the simple form of (\ref{eq:gam-surv-var-alt}).

\vspace{-.2in}
\subsection{Hazard odds model}

First, we derive model-robust variance estimation for $ \log \hat Q_k(x_0)$, where $\hat Q_k( x_0) = \prod_{j=1}^k \{1- \hat q_j(x_0)\}$ from (\ref{eq:bet-surv}).
Consider the Taylor expansion
\begin{align*}
& \log \hat Q_k(x_0) - \log \bar Q_k(x_0) = \sum_{j=1}^k \left\{ \log \left( 1-\frac{ \me^{\hat\beta_{0j}}}{1+ \me^{\hat\beta_{0j}}} \right) -
\log \left(1- \frac{ \me^{\bar\beta_{0j}}}{1+ \me^{\bar\beta_{0j}}} \right) \right\} \\
& = - \sum_{j=1}^k (1+\me^{\bar\beta_{0j}}) \left(\frac{ \me^{\hat\beta_{0j}}}{1+ \me^{\hat\beta_{0j}}} -\frac{ \me^{\bar\beta_{0j}}}{1+ \me^{\bar\beta_{0j}}}\right) + o_p(n^{-1/2}),
\end{align*}
where $\bar Q_k(x_0) = \prod_{j=1}^k \{ 1 /(1+\me^{\bar\beta_{0j}}) \}$ and
$ \me^{\bar\beta_{0j}} = E (\tilde R_j \tilde D_j) /E(\tilde R_j (1-\tilde D_j)\me^{\tilde X^\T \bar\beta})$, which is the probability limit of $\me^{\hat\beta_{0j}}$. Then we use the decomposition
\begin{align*}
& \frac{ \me^{\hat\beta_{0j}}}{1+ \me^{\hat\beta_{0j}}} -\frac{ \me^{\bar\beta_{0j}}}{1+ \me^{\bar\beta_{0j}}} =
\left\{ \frac{\sum_{i=1}^n R_{ji} D_{ji}} {\sum_{i=1}^n R_{ji} D_{ji}+R_{ji}(1-D_{ji})\me^{ X_i^\T \bar\beta} } - \frac{ \me^{\bar\beta_{0j}}}{1+ \me^{\bar\beta_{0j}}}\right\} \\
& \quad + \left\{ \frac{\sum_{i=1}^n R_{ji} D_{ji}} {\sum_{i=1}^n R_{ji} D_{ji}+R_{ji}(1-D_{ji}) \me^{ X_i^\T \hat\beta}} -
\frac{\sum_{i=1}^n R_{ji} D_{ji}} {\sum_{i=1}^n R_{ji} D_{ji}+R_{ji}(1-D_{ji}) \me^{ X_i^\T \bar\beta}}  \right\}.
\end{align*}
The first term can be approximated as
\begin{align*}
& \frac{\sum_{i=1}^n R_{ji} D_{ji}} {\sum_{i=1}^n R_{ji} D_{ji}+R_{ji}(1-D_{ji})\me^{ X_i^\T \bar\beta}} -\frac{ \me^{\bar\beta_{0j}}}{1+ \me^{\bar\beta_{0j}}} \\
& = \frac{ n^{-1} \sum_{i=1}^n R_{ji}D_{ji} E\{\tilde R_j (1-\tilde D_j)\me^{\tilde X^\T \bar\beta}\}- R_{ji}(1-D_{ji})\me^{X^\T_i \bar\beta} E (\tilde R_j \tilde D_j) }
{E^2 (\tilde R_j \tilde D_j + \tilde R_j (1-\tilde D_j) \me^{\tilde X^\T \bar\beta} ) } + o_p(n^{-1/2}) . 
\end{align*}
The second term can be approximated as
\begin{align*}
&  \frac{\sum_{i=1}^n R_{ji} D_{ji}} {\sum_{i=1}^n R_{ji} D_{ji}+R_{ji}(1-D_{ji}) \me^{ X_i^\T \hat\beta}} -
\frac{\sum_{i=1}^n R_{ji} D_{ji}} {\sum_{i=1}^n R_{ji} D_{ji}+R_{ji}(1-D_{ji}) \me^{ X_i^\T \bar\beta}}  \\
& = \frac{- E(\tilde R_j \tilde D_j) E(\tilde R_j (1-\tilde D_j) \me^{\tilde X^\T \bar\beta} \tilde X^\T) } {E^2(\tilde R_j \tilde D_j + \tilde R_j (1-\tilde D_j) \me^{\tilde X^\T \bar\beta} ) }  (\hat\beta-\bar\beta) + o_p(n^{-1/2}) \\
& =  \frac{- E(\tilde R_j \tilde D_j) E(\tilde R_j (1-\tilde D_j) \me^{\tilde X^\T \bar\beta} \tilde X^\T) } {E^2(\tilde R_j \tilde D_j + \tilde R_j (1-\tilde D_j) \me^{\tilde X^\T \bar\beta} ) }  H^{-1}(\bar\beta) \frac{1}{n} \sum_{i=1}^n g_\bullet (Y_i, \delta_i, X_i; \bar\beta)+ o_p(n^{-1/2}),
\end{align*}
where $g_\bullet (Y,\delta,X;\beta) =\sum_{j=1}^J g_j(Y,\delta,X;\beta)$.
Combining the preceding four displays yields the asymptotic expansion
\begin{align}
\log \hat Q_k(x_0) - \log \bar Q_k(x_0) = \frac{1}{n} \sum_{i=1}^n \psi_k (Y_i,\delta_i,X_i; \bar\beta) + o_p(n^{-1/2}), \label{eq:bet-surv-expan}
\end{align}
where
\begin{align*}
& \psi_k(Y,\delta,X; \beta) \\
& =- \sum_{j=1}^k (1+\me^{\beta_{0j}}) \left\{ \frac{R_j D_j E\{\tilde R_j (1-\tilde D_j)\me^{\tilde X^\T \bar\beta}\}- R_j(1-D_j)\me^{X^\T \bar\beta} E (\tilde R_j \tilde D_j) }
{E^2 (\tilde R_j \tilde D_j + \tilde R_j (1-\tilde D_j) \me^{\tilde X^\T \bar\beta} ) } \right. \\
& \qquad \left. -
\frac{E(\tilde R_j \tilde D_j) E(\tilde R_j (1-\tilde D_j) \me^{\tilde X^\T \bar\beta} \tilde X^\T) } {E^2(\tilde R_j \tilde D_j + \tilde R_j (1-\tilde D_j) \me^{\tilde X^\T \bar\beta} ) }
H^{-1}(\beta)g_\bullet (Y, \delta, X; \beta)\right\} .
\end{align*}
Then a model-robust variance estimator for $\log \hat Q_k(x_0)$ is  $n^{-2} \sum_{i=1}^n  \hat\psi_k^2 (Y_i,\delta_i,X_i; \hat\beta)$, where
$\hat \psi_k (Y,\delta,X; \beta)$ is defined as $\psi_k (Y,\delta, X;\beta)$ with
$B(\beta)$ replaced by $\hat B(\beta)$,
$g_\bullet (Y, \delta, X; \beta)$ replaced by $\hat g_\bullet (Y, \delta, X;\beta) =\sum_{j=1}^J \hat g_j(Y,\delta,X;\beta)$, and
$E (\tilde R_j \tilde D_j)$, $E(\tilde R_j (1-\tilde D_j) \me^{ \tilde X^\T \beta})$, and $E( \tilde R_j (1-\tilde D_j) \me^{\tilde X^\T \beta} \tilde X )$ replaced by the corresponding sample averages.

For model-based variance estimation, suppose that model (\ref{eq:odds-model2}) is correctly specified and hence $\bar\beta=\beta^*$.
Then $g_j(Y,\delta,X;\beta)$ reduces to $g_{j1}(Y,\delta,X;\beta)$ only,
and $g_\bullet(Y,\delta,X;\beta)$ reduces to $g_{1\bullet} (Y,\delta,X;\beta)$ defined as $\sum_{j=1}^J g_{j1}(Y,\delta,X;\beta)$.
We return to the asymptotic expansion (\ref{eq:bet-surv-expan}), and use the fact that
the individual terms $R_j D_j E\{\tilde R_j (1-\tilde D_j)\me^{\tilde X^\T \bar\beta}\}- R_j(1-D_j)\me^{X^\T \bar\beta} E (\tilde R_j \tilde D_j) $ are uncorrelated not only with each other for $j=1,\ldots,J$
but also with $g_{1\bullet}(Y, \delta, X; \beta^*)$. By (\ref{eq:bet-G-calc}) in the supplement, the asymptotic variance of $\log \hat Q_k(x_0)$ can be simplified as
\begin{align}
\frac{1}{n} \left[ \sum_{j=1}^k  \frac{ (1+\me^{\beta^*_{0j}})^2 E (R_j \me^{\beta^*_{0j} + X^\T \beta^* })
E^2 (\tilde R_j(1-\tilde D_j) \me^{\tilde X^\T \beta^*}) }{ E^4 (\tilde R_j \tilde D_j + \tilde R_j (1-\tilde D_j)\me^{ \tilde X^\T \beta^*})}
+ \Gamma_k^\T \Sigma  \Gamma_k \right], \label{eq:bet-surv-var}
\end{align}
where  $\me^{\beta^*_{0j}}=
E( \tilde R_j\tilde D_j) /E(\tilde R_j(1-\tilde D_j) \me^{ \tilde X^\T \beta^*})$, and
\begin{align*}
\Gamma_k = \sum_{j=1}^k (1+\me^{\beta^*_{0j}}) \frac{E(\tilde R_j \tilde D_j) E(\tilde R_j(1-\tilde D_j) \me^{\tilde X^\T \beta^*} \tilde X) }
{E^2 (\tilde R_j \tilde D_j + \tilde R_j (1-\tilde D_j)\me^{ \tilde X^\T \beta^*})} .
\end{align*}
A model-based variance estimator for $\log \hat Q_k(x_0)$ is obtained from (\ref{eq:bet-surv-var}) with $\beta^*$ replaced by $\hat\beta$, $\Sigma$ replaced by
$\hat \Sigma_{\text{b}}$ or $\hat \Sigma_{\text{b2}}$, and
all expectations replaced by the corresponding sample averages, for example, the expectation $E (R_j \me^{\beta^*_{0j} + X^\T \beta^* })$ replaced by
$n^{-1} \sum_{i=1}^n R_j \me^{\hat\beta_{0j}+ X_i^\T \hat\beta}$,
where $\me^{\hat\beta_{0j}} = ( \sum_{l=1}^n R_{jl} D_{jl}) / \{\sum_{l=1}^n R_{jl}(1-D_{jl}) \me^{X_l^\T \hat\beta}\}$.

\vspace{.3in}
\centerline{\bf\Large References}

\begin{description}\addtolength{\itemsep}{-.1in}
\item Allison, P.D. (1982) Discrete-time methods for the analysis of event histories, {\em Sociological Methodology}, 13, 61--98.

\item Andersen, P.K., Borgan, O., Gill, R.D., and Keiding, N. (1993) {\em Statistical Models Based on Counting Processes}, New York: Springer.

\item Breslow, N.E. (1974) Covariance analysis of censored survival data, {\em Biometrics},  30, 89--100.

\item Breslow, N.E. (1981) Odds ratio estimators when the data are sparse, {\em Biometrika}, 68, 73--84.

\item Buja, A., Berk, R., Brown, L., George, E., Pitkin, E., Traskin, M., Zhao, L., and Zhang, K. (2019) Models as approximations I:
Consequences illustrated with linear regression, {\em Statistical Science}, 34, 523-544.

\item Cochran, W.G. (1954) Some methods for strengthening the common $\chi^2$ tests, {\em Biometrics}, 10, 417--451.

\item Cox, D.R. (1972) Regression models and life tables (with discussion), {\em Journal of the Royal Statistical Society}, Ser.~B, 34, 187--220.

\item Cox, D.R. and Oaks, D.O. (1984) {\em Analysis of Survival Data}, London: Chapman \& Hall.

\item Efron, B. (1977) The efficiency of Cox's likelihood function for censored data, {\em  Journal of the American Statistical Association}, 72, 557--565.


\item Kalbfleisch, J.D. and Prentice, R.L. (1980) {\em The Statistical Analysis of Failure Time Data}, New York: Wiley.

\item Kaplan, E.L. and Meier, P. (1958) Nonparametric estimation from incomplete observations, {\em Journal of the American Statistical Association}, 53, 457--481.

\item Lin, D.Y. and Wei, L.J. (1989) The robust inference for the Cox proportional hazards model, {\em Journal of the American Statistical Association}, 84, 1074--1079.

\item Lindsay, B.G. (1980) Nuisance parameters, mixture models and the efficiency of partial likelihood estimators, {\em Philosophical Transactions of the Royal Society}, Ser.~A, 296, 639--665.

\item Lindsay, B.G. (1983) Efficiency of the conditional score in a mixture setting, {\em Annals of Statistics}, 11, 486--197.

\item Manski, C.F. (1988) {\em Analog Estimation Methods in Econometrics}. New York: Chapman \& Hall.

\item Mantel, N. and Haenszel, W.M. (1959)
Statistical aspects of the analysis of data from retrospective studies of disease, {\em Journal of the National Cancer Institute}, 22, 719--748.

\item Peto, R. (1972) Contribution to the discussion of Cox (1972): Regression models and life tables, {\em Journal of the Royal Statistical Society}, Ser.~B, 34, 205--207.

\item Prentice, R.L. and Gloeckler, L.A. (1978) Regression analysis of grouped survival data with application to breast cancer data, {\em Biometrics}, 34, 57--67.

\item Robins, J.M., Breslow, N.E., Greenland, S. (1986) Estimators of the Mantel--Haenszel variance consistent in both
sparse data and large strata limiting models, {\em Biometrics}, 42, 311--324.

\item Tan, Z. (2019) Analysis of odds, probability, and hazard ratios: From 2 by 2 ables to two-sample survival data, arXiv:1911.10682.

\item Tan, Z. (2020a) Regularized calibrated estimation of propensity scores with model misspecification and high-dimensional data, {\em Biometrika}, 107, 137--158.

\item Tan, Z. (2020b) {\em dSurvival: Discrete-time Survival Analysis}, R package version 1.0, available at \url{http://www.stat.rutgers.edu/~ztan}.

\item Therneau, T.M. (2015) {\em A Package for Survival Analysis}, version 2.38.

\item Therneau, T.M., Grambsch, P.M., and Fleming, T.R. (1990) Martingale based residuals for survival models, {\em Biometrika}, 77, 147--160.

\item Therneau, T.M. and Grambsch, P.M. (2000) {\em Modeling Survival Data: Extending the Cox Model}, New York: Springer.

\item Thompson, W.A.Jr. (1977) On the treatment of grouped observations in life studies, {\em Biometrics}, 33, 463--470.

\item Tsiatis, A.A. (1981) A large sample study of Cox's regression model, {\em Annals of Statistics}, 9, 93--108.

\item White, H. (1982) Maximum likelihood estimation of misspecified models, {\em Econometrica}, 50, 1--25.

\item Willett, J.B. and Singer, J.D. (2004) Discrete-time survival analysis, in {\em SAGE Handbook of Quantitative Methodology for the Social Sciences}, ed.~Kaplan, D., 200--213.

\end{description}

\clearpage

\setcounter{page}{1}

\setcounter{section}{0}
\setcounter{equation}{0}

\setcounter{figure}{0}
\setcounter{table}{0}

\renewcommand{\theequation}{S\arabic{equation}}
\renewcommand{\thesection}{\Roman{section}}

\renewcommand\thefigure{S\arabic{figure}}
\renewcommand\thetable{S\arabic{table}}

\begin{center}
{\Large Supplementary Material for}

{\Large ``Consistent and robust inference in hazard probability and odds models with discrete-time survival data"}

\vspace{.1in} {\large Zhiqiang Tan}
\end{center}
\vspace{.1in}

\section{Relationship with two-sample survival analysis}

We discuss how the point and variance estimators proposed in regression models with survival data are related to
those in Tan (2019) for two-sample survival analysis.

Suppose that the covariate $X$ is binary, taking values 1 or 0, corresponding to the first or second group.
We use the notation similarly as in Tan (2019, Section 3).
For $j=1,\ldots, J$, denote by $P_{1j} = P( R_j =1, X=1)$ and $P_{2j} = P(R_j=1, X=0)$ the probabilities of being included in the $j$th risk set from the first and second groups,
and by $p_{11j} = P( D_j=1 | R_j=1, X=1) = 1-p_{12j}$, and $p_{21j} = P(D_j = 1 | R_j=1, X=0) = 1-p_{22j}$ the hazard probabilities in the first and second groups.
Denote by $(n_{1j},n_{2j})$ the sizes of the first and second groups,
by $(n_{11j}, n_{21j})$ the number of events in the first and second groups in the $j$th risk set.
Denote $n_{12j} = n_{1j} - n_{11j}$ and $n_{22j} = n_{2j} - n_{21j}$.

\subsection{Hazard probability model}

In a two-sample setting, the hazard probability model (\ref{eq:prob-model}) can be stated as
\begin{align*}
 p_{11j}  = \exp (\gamma^*) p_{21j} , \quad j=1,\ldots,J. 
\end{align*}
The point estimator $\hat\gamma$ in Tan (2019) is defined as a solution to
\begin{align*}
0 = \sum_{j=1}^J \frac{n_{11j} n_{2j} - \me^\gamma n_{21j} n_{1j} } { n_{1j} \me^\gamma + n_{2j}} , 
\end{align*}
which is shown to coincide with the Breslow--Peto estimating equation.
Moreover, the model-robust variance estimator for $\hat\gamma$ in Tan (2019) is also shown to be identical to that in Lin \& Wei (1989)
extended for the Breslow--Peto estimator with  tied events, and hence it is identical to the proposed variance estimator $\hat V_{\text{r}}$ in Section~\ref{sec:prob-model}.

We show that the proposed variance estimator $\hat V_{\text{b2}}$ reduces to a model-based variance estimator for $\hat\gamma$ in Tan (2019, Section 2.2.2).
In fact, direct calculation yields
\begin{align*}
& \hat B(\gamma) = \frac{1}{n} \sum_{j=1}^J (n_{11j} + n_{21j}) \times \\
& \quad \left\{ \frac{n_{1j} \me^\gamma}{n_{1j}\me^\gamma+n_{2j}} \left(1- \frac{n_{1j}\me^\gamma}{n_{1j}\me^\gamma+n_{2j}} \right)^2
+  \frac{ n_{2j} }{n_{1j}\me^\gamma+n_{2j}} \left(0- \frac{n_{1j} \me^\gamma }{n_{1j}\me^\gamma+n_{2j}} \right)^2  \right\}  \\
& = \sum_{j=1}^J (n_{11j} + n_{21j})  \frac{n_{1j} n_{2j} \me^\gamma}{(n_{1j}\me^\gamma+n_{2j})^2 } ,
\end{align*}
and
\begin{align*}
& \hat A_{\text{b2}} (\gamma) = \frac{1}{n} \sum_{j=1}^J \left\{ n_{12j}\me^\gamma \frac{n_{2j} n_{21j}}{(n_{1j}\me^\gamma+n_{2j})^2 } +
n_{22j} \frac{n_{1j}\me^\gamma n_{11j}}{( n_{1j}\me^\gamma+n_{2j})^2 }  \right\} \\
& = \frac{1}{n} \sum_{j=1}^J  \me^\gamma \frac{n_{2j} n_{12j} n_{21j} + n_{1j}  n_{11j}n_{22j} }{(n_{1j}\me^\gamma+n_{2j})^2 }.
\end{align*}
Then $\hat V_{\text{b2}} = \hat B^{-1}(\hat\gamma) \hat A_{\text{b2}}(\hat\gamma) \hat B^{-1}( \hat\gamma)$ reduces to the model-based variance estimator in Tan (2019, Section 2.2.2)
after proper rescaling of the sample size.

\subsection{Hazard odds model}

In a two-sample setting, the hazard odds model (\ref{eq:odds-model}) can be stated as
\begin{align*}
\frac{p_{11j}}{1-p_{11j}} = \me^{\beta^*} \frac{p_{21j}}{1-p_{21j}}, \quad j=1,\ldots,J. 
\end{align*}
The point estimator $\hat\beta$ in Tan (2019) is defined as a solution to
\begin{align*}
0 = \sum_{j=1}^J \frac{n_{11j} n_{22j} - \me^\beta n_{12j} n_{21j} } { n_{1j} \me^\beta + n_{2j}} ,
\end{align*}
which is a weighted extension of the Mantel--Haenszel estimator in $2\times 2$ tables.
The proposed estimating equation (\ref{eq:bet-est}) can be directly calculated as
\begin{align*}
0 & = \sum_{j=1}^J \frac{ n_{11j} (n_{12j} \me^\beta + n_{22j}) - n_{12j} \me^\beta ( n_{11j} + n_{21j} ) }{ n_{1j}\me^\beta+n_{2j} } \\
& = \sum_{j=1}^J \frac{n_{11j} n_{22j} - \me^\beta n_{12j} n_{21j} } { n_{1j} \me^\beta + n_{2j}} ,
\end{align*}
hence the same as the Mantel--Haenszel estimating equation in Tan (2019).

We show that the variance estimator $\hat \Sigma_{\text{b3}}$ reduces to a model-based variance estimator for $\hat\beta$ in Tan (2019, Section 2.1.2),
i.e., the variance estimator in Robins et al.~(1986) in the setting of common odds ratios. In fact, by direct calculation,
\begin{align*}
& \hat H(\beta) =  \frac{1}{n} \sum_{j=1}^J \left\{  \frac{n_{12j}\me^\beta n_{21j}}{{ n_{1j}\me^\beta+n_{2j}}}  \left(1-\frac{n_{1j}\me^\beta}{n_{1j}\me^\beta+n_{2j}} \right) +
 \frac{(-n_{22j}) n_{11j}}{{ n_{1j}\me^\beta+n_{2j}}}  \left( 0 -\frac{n_{1j}\me^\beta}{n_{1j}\me^\beta+n_{2j}} \right) \right\} \\
& =  \frac{1}{n} \sum_{j=1}^J \me^\beta \frac{ n_{1j} n_{11j} n_{22j}  + n_{2j} n_{12j} n_{21j} }{ ( n_{1j}\me^\beta+n_{2j})^2 },
\end{align*}
and
\begin{align*}
& \hat G_{\text{b3}} (\beta) =  \frac{1}{n} \sum_{j=1}^J
\frac{ n_{12j} \me^\beta (n_{22j} + n_{21j}\me^\beta) n_{21j} + n_{22j} (n_{12j} \me^\beta + n_{11j}) n_{11j} } {( n_{1j}\me^\beta+n_{2j})^2  } \\
& =  \frac{1}{n} \sum_{j=1}^J
\frac{ \me^\beta n_{12j} n_{21j} (n_{22j}+ \me^\beta n_{21j}) +  n_{11j} n_{22j} (n_{11j} + \me^\beta n_{12j}) } {( n_{1j}\me^\beta+n_{2j})^2  }
\end{align*}
Then $\hat \Sigma_{\text{b3}} = \hat H^{-1}(\hat\beta) \hat G_{\text{b3}}(\hat\beta) \hat H^{-1}( \hat\beta)$ reduces to the model-based variance estimator in Tan (2019, Section 2.1.2)
after proper rescaling of the sample size.

The variance estimator $\hat \Sigma_{\text{b2}}$ is defined as  $\hat H^{-1}(\hat\beta) \hat G_{\text{b2}}(\hat\beta) \hat H^{-1}( \hat\beta)$, where $\hat G_{\text{b2}} (\beta)$ can be calculated as
\begin{align*}
& \hat G_{\text{b2}} (\beta) =  \frac{1}{n} \sum_{j=1}^J
\frac{ (n_{12j} \me^\beta n_{21j} + n_{22j} n_{11j} \me^\beta ) +
( n_{1j} \me^\beta n_{22j} n_{21j} + n_{2j} n_{12j} \me^\beta n_{11j} ) } {( n_{1j}\me^\beta+n_{2j})^2 } \\
& =\frac{1}{n} \sum_{j=1}^J \me^\beta \frac{ n_{12j} n_{21j} + n_{22j} n_{11j} +
 n_{1j} n_{22j} n_{21j} + n_{2j} n_{12j} n_{11j} } {( n_{1j}\me^\beta+n_{2j})^2 } .
\end{align*}
This differs from $\hat G_{\text{b3}}(\beta)$ based on Robins et al.~(1986) as well as another choice based on Flander (1985), which is stated in Tan (2019, Supplement).
In the special case of only one event at time $t_j$, for example, $n_{11j}=0$, $n_{1j}=n_{12j}$, $n_{21j}=1$, and $n_{2j}=n_{22j}+1$,
the $j$th terms in $\hat H(\beta)$ and $\hat G_{\text{b2}}(\beta)$ both reduce to $ n_{1j} (n_{22j}+1)$.

Finally, we show that the model-robust variance estimator $\hat\Sigma_{\text{r}}$ is identical to that in Tan (2019).
It suffices to verify that $g_j(Y,\delta,X; \beta)$ coincides with the version, denoted as $g_j^{\text{TS}}(Y,\delta,X; \beta)$, in Tan (2019, Proposition 7).
Denote  $I_{1j} = 1\{ R_j=1, X=1\}$ and $I_{2j} = 1\{ R_j=1, X=0\}$ associated with $(P_{1j}, P_{2j})$, and $I_{11j} = 1\{ D_j=1, R_j=1, X=1\}$ and $I_{21j} = 1\{ D_j=1, R_j=1, X=0\}$
associated with $P_{11j} = P_{1j} p_{11j}$ and $P_{21j} = P_{2j} p_{21j}$, etc, similarly as in Tan (2019).
Direct calculation yields
\begin{align*}
& g_j(Y,\delta,X; \beta) \\
& = \frac{ I_{11j} ( P_{12j} \me^\beta + P_{22j} ) - I_{12j} \me^\beta ( P_{11j} + P_{21j})} { P_{1j} \me^\beta + P_{2j}} \left( 1- \frac{P_{12j} \me^\beta}{P_{12j} \me^\beta + P_{22j}} \right) \\
& \quad + \frac{ I_{21j} ( P_{12j} \me^\beta + P_{22j} ) - I_{22j} \me^\beta ( P_{11j} + P_{21j})} { P_{1j} \me^\beta + P_{2j}} \left( 0- \frac{P_{12j} \me^\beta}{P_{12j} \me^\beta + P_{22j}} \right) \\
& \quad - \frac{ P_{11j} P_{22j} -  \me^\beta P_{12j} P_{21j} } { P_{1j} \me^\beta + P_{2j}}
\left( \frac{I_{1j}\me^\beta +I_{2j}}{P_{1j} \me^\beta + P_{2j}} - \frac{I_{12j} \me^\beta + I_{22j}}{P_{12j} \me^\beta + P_{22j}} \right) ,
\end{align*}
and
\begin{align*}
& g_j^{\text{TS}}(Y,\delta,X; \beta) \\
& = \frac{P_{22j} + \me^\beta P_{21j} } { P_{1j} \me^\beta + P_{2j}} \left( I_{11j} - \frac{P_{11j}}{P_{1j}} I_{1j} \right)
+  \frac{P_{11j} + \me^\beta P_{12j} } { P_{1j} \me^\beta + P_{2j}} \left( I_{21j} - \frac{P_{21j}}{P_{2j}} I_{2j} \right) \\
& \quad + \frac{ P_{11j} P_{22j} -  \me^\beta P_{12j} P_{21j} } { (P_{1j} \me^\beta + P_{2j} )^2 }
\left(\frac{P_{2j}}{P_{1j}} I_{1j} + \frac{\me^\beta P_{1j}}{P_{2j}} I_{2j} \right) \\
& =  \frac{P_{22j} + \me^\beta P_{21j} } { P_{1j} \me^\beta + P_{2j}}  I_{11j}
+  \frac{P_{11j} + \me^\beta P_{12j} } { P_{1j} \me^\beta + P_{2j}} I_{21j} \\
& \quad -\frac{ \me^{2\beta} P_{21j} P_{11j}  + \me^\beta P_{22j} P_{11j} + \me^\beta P_{21j} P_{2j} } { (P_{1j} \me^\beta + P_{2j} )^2 } I_{1j}
 + \frac{ P_{11j} P_{21j}  + \me^\beta P_{12j} P_{21j} + \me^\beta P_{11j} P_{1j} } { (P_{1j} \me^\beta + P_{2j} )^2 } I_{2j} .
\end{align*}
Treating $I_{12j} = I_{1j} - I_{11j}$, the coefficient of $I_{11j}$ in $g_j(Y,\delta,X; \beta)$ can be shown to be identical to that in $g_j^{\text{TS}}(Y,\delta,X; \beta)$:
\begin{align*}
& \frac{P_{12j} \me^\beta + P_{22j}} { P_{1j} \me^\beta + P_{2j}} \frac{P_{22j}}{P_{12j} \me^\beta + P_{22j}}  -
\frac{ P_{11j} P_{22j} -  \me^\beta P_{12j} P_{21j} } { P_{1j} \me^\beta + P_{2j}} \frac{\me^\beta}{P_{12j} \me^\beta + P_{22j}} \\
& = \frac{P_{22j} + \me^\beta P_{21j} } { P_{1j} \me^\beta + P_{2j}} .
\end{align*}
Moreover, the coefficient of $I_{1j}$ in $g_j(Y,\delta,X; \beta)$ can be shown to be identical to that in $g_j^{\text{TS}}(Y,\delta,X; \beta)$:
\begin{align*}
& -\frac{\me^\beta ( P_{11j}+P_{22j}) } { P_{1j} \me^\beta + P_{2j}} \frac{P_{22j}}{P_{12j} \me^\beta + P_{22j}} -
\frac{ P_{11j} P_{22j} -  \me^\beta P_{12j} P_{21j} } { P_{1j} \me^\beta + P_{2j}} \left( \frac{\me^\beta}{P_{1j} \me^\beta + P_{2j}}  - \frac{\me^\beta}{P_{12j} \me^\beta + P_{22j}} \right) \\
& = - \frac{\me^\beta P_{21j}}{ P_{1j} \me^\beta + P_{2j}} - \frac{ P_{11j} P_{22j} -  \me^\beta P_{12j} P_{21j} } { P_{1j} \me^\beta + P_{2j}} \frac{\me^\beta}{P_{1j} \me^\beta + P_{2j}} \\
& = -\frac{ \me^{2\beta} P_{21j} P_{11j}  + \me^\beta P_{22j} P_{11j} + \me^\beta P_{21j} P_{2j} } { (P_{1j} \me^\beta + P_{2j} )^2 }.
\end{align*}
Similarly, the coefficients of $I_{21j}$ and $I_{2j}$ in $g_j(Y,\delta,X; \beta)$ can be shown to be identical to those in $g_j^{\text{TS}}(Y,\delta,X; \beta)$.

\section{Technical details}

\noindent\textbf{Proof of Proposition~\ref{pro:gam-model-robust}}.\;
It suffices to show that for $j=1,\ldots,J$,
\begin{align*}
& \hat\zeta_j(\bar\gamma) = \frac{1}{n} \sum_{i=1}^n h_j (Y_i,\delta_i,X_i ; \bar\gamma) + o_p(n^{-1/2}),
\end{align*}
that is,
\begin{align}
&  \frac{1}{n} \sum_{i=1}^n R_{ji} D_{ji}  X_i -
 \frac{\sum_{l=1}^n R_{jl} D_{jl} } {\sum_{l=1}^n R_{jl} \me^{ X_l^\T \bar\gamma}}  \frac{1}{n} \sum_{i=1}^n R_{ji}  \me^{X_i^\T \bar \gamma}  X_i \nonumber \\
& = \frac{1}{n} \sum_{i=1}^n  R_{ji} \left\{ D_{ji} - \frac{E( \tilde R_j\tilde D_j)} {E(\tilde R_j \me^{ \tilde X^\T \bar\gamma})} \me^{X_i^\T \bar\gamma} \right\}
 \left\{ X_i - \frac{E( \tilde R_j \me^{\tilde X^\T \bar\gamma} \tilde X )} {E(\tilde R_j \me^{ \tilde X^\T \bar\gamma})}  \right\}   + o_p(n^{-1/2}). \label{eq:prf1-gam-model-robust}
\end{align}
Substituting the expansion (\ref{eq:gam-surv-first-term}) into the left hand side of (\ref{eq:prf1-gam-model-robust}) yields
\begin{align}
&  \frac{1}{n} \sum_{i=1}^n R_{ji} D_{ji}  X_i -
 \frac{\sum_{l=1}^n R_{jl} D_{jl} } {\sum_{l=1}^n R_{jl} \me^{ X_l^\T \bar\gamma}}  \frac{1}{n} \sum_{i=1}^n R_{ji}  \me^{X_i^\T \bar \gamma}  X_i \nonumber \\
& = \frac{1}{n} \sum_{i=1}^n R_{ji} D_{ji}  X_i - \frac{E( \tilde R_j\tilde D_j)} {E(\tilde R_j \me^{ \tilde X^\T \bar\gamma})}
\frac{1}{n} \sum_{i=1}^n R_{ji}  \me^{X_i^\T \bar \gamma}  X_i  \nonumber \\
& \quad -\left\{\frac{ n^{-1} \sum_{i=1}^n R_{ji} ( D_{ji} - \me^{\bar\gamma_{0j} + X^\T_i \bar\gamma} ) } {E (\tilde R_j \me^{\tilde X^\T \bar\gamma} ) }  \right\}
\frac{1}{n} \sum_{i=1}^n R_{ji}  \me^{X_i^\T \bar \gamma}  X_i  + o_p(n^{-1/2}) ,   \label{eq:prf2-gam-model-robust}
\end{align}
where $ \me^{\bar\gamma_{0j}} = E (\tilde R_j \tilde D_j) /E( \tilde R_j \me^{\tilde X^\T \bar\gamma})$.
The sum of the first two terms on the right hand side of (\ref{eq:prf2-gam-model-robust}) gives
\begin{align*}
& \frac{1}{n} \sum_{i=1}^n R_{ji} D_{ji}  X_i - \frac{E( \tilde R_j\tilde D_j)} {E(\tilde R_j \me^{ \tilde X^\T \bar\gamma})}
\frac{1}{n} \sum_{i=1}^n R_{ji}  \me^{X_i^\T \bar \gamma}  X_i
 = \frac{1}{n} \sum_{i=1}^n   R_{ji} \left\{ D_{ji} - \frac{E( \tilde R_j\tilde D_j)} {E(\tilde R_j \me^{ \tilde X^\T \bar\gamma})} \me^{X_i^\T \bar\gamma} \right\}  X_i .
\end{align*}
The third term on the right hand side of (\ref{eq:prf2-gam-model-robust}) can be approximated as
\begin{align*}
& \left\{\frac{ n^{-1} \sum_{i=1}^n R_{ji} ( D_{ji} - \me^{\bar\gamma_{0j} + X^\T_i \bar\gamma} ) } {E (\tilde R_j \me^{\tilde X^\T \bar\gamma} ) }  \right\}
\frac{1}{n} \sum_{i=1}^n R_{ji}  \me^{X_i^\T \bar \gamma}  X_i \\
& = \frac{E( \tilde R_j \me^{\tilde X^\T \bar\gamma} \tilde X ) } {E (\tilde R_j \me^{\tilde X^\T \bar\gamma} ) }
\frac{1}{n} \sum_{i=1}^n R_{ji} ( D_{ji} - \me^{\bar\gamma_{0j} + X^\T_i \bar\gamma} ) + o_p(n^{-1/2}).
\end{align*}
Combining the preceding two displays gives the right hand side of (\ref{eq:prf1-gam-model-robust}).
{\hfill $\Box$ \vspace{.1in}}

\noindent\textbf{Proof of Proposition~\ref{pro:gam-model-based}}.

(i) Suppose that model (\ref{eq:prob-model2}) is correctly specified. We show that
$E \{\hat v_j(\gamma^*) | R_{j,1:n}, X_{1:n} \} = v_j (\gamma^*)$ for $j=1,\ldots,J$,
where $v_j(\gamma^*) = n \,\var \{ \hat\zeta_j(\gamma^*)  | R_{j,1:n}, X_{1:n} \}$ and
\begin{align*}
\hat\zeta_j(\gamma^*) & = \frac{1}{n} \sum_{i=1}^n R_{ji} \left( D_{ji} - \frac{\sum_{l=1}^n R_{jl} D_{jl} } {\sum_{l=1}^n R_{jl} \me^{ X_l^\T \gamma^*}} \me^{X_i^\T \gamma^*} \right)  X_i \\
& = \frac{1}{n} \sum_{i=1}^n R_{ji} D_{ji}  \left( X_i - \frac{\sum_{l=1}^n R_{jl}\me^{X_l^\T \gamma^*} X_l } {\sum_{l=1}^n R_{jl} \me^{ X_l^\T \gamma^*}}  \right) .
\end{align*}
First, we see that under model (\ref{eq:prob-model2}),
\begin{align*}
& E\{ \hat\zeta_j(\gamma^*) | R_{j1}, \ldots, R_{jn}, X_1,\ldots,X_n \} \\
& =  \frac{1}{n} \sum_{i=1}^n R_{ji} p_j(X_i)  \left( X_i - \frac{\sum_{l=1}^n R_{jl}\me^{X_l^\T \gamma^*} X_l } {\sum_{l=1}^n R_{jl} \me^{ X_l^\T \gamma^*}}  \right)\\
& =  \frac{1}{n} \sum_{i=1}^n R_{ji} \left( p_j(X_i) - \frac{\sum_{l=1}^n R_{jl} p_j(X_l) } {\sum_{l=1}^n R_{jl} \me^{ X_l^\T \gamma^*}} \me^{X_i^\T \gamma^*} \right)  X_i = 0,
\end{align*}
where $p_j(X) = \me^{\gamma^*_{0j} + X^\T \gamma^*}$.
Moreover,
\begin{align*}
& n\,\var\{ \hat\zeta_j(\gamma^*) | R_{j1}, \ldots, R_{jn}, X_1,\ldots,X_n \} \\
& = \frac{1}{n} \sum_{i=1}^n R_{ji} p_j(X_i) (1-p_j(X_i))  \left( X_i - \frac{\sum_{l=1}^n R_{jl}\me^{X_l^\T \gamma^*} X_l } {\sum_{l=1}^n R_{jl} \me^{ X_l^\T \gamma^*}}  \right)^{\otimes 2} \\
& = \frac{n^{-1}}{ (\sum_{l=1}^n R_{jl} \me^{ X_l^\T \gamma^*} )^2} \sum_{i=1}^n R_{ji} p_j(X_i) (1-p_j(X_i))\left\{ \sum_{l=1}^n R_{jl}\me^{X_l^\T \gamma^*} (X_i-X_l) \right\}^{\otimes 2}
\end{align*}
Next, we calculate
\begin{align*}
& E \{ \hat v_j(\gamma^*) |  R_{j1}, \ldots, R_{jn}, X_1,\ldots,X_n \} \\
& = \frac{n^{-1}}{(\sum_{l=1}^n R_{jl} \me^{X_l^\T \gamma^*} )^2}  \\
& \quad \times \sum_{i=1}^n R_{ji} (1-p_j(X_i)) \me^{X_i^\T \gamma^*} \sum_{l=1}^n R_{jl} \me^{X_l^\T \gamma^*} (X_i-X_l) \sum_{k=1}^n R_{jk} p_j(X_k) (X_i-X_k)^\T  \\
& = \frac{n^{-1}}{(\sum_{l=1}^n R_{jl} \me^{X_l^\T \gamma^*} )^2}  \\
& \quad \times \sum_{i=1}^n R_{ji} p_j(X_i) (1-p_j(X_i))\sum_{l=1}^n R_{jl} \me^{X_l^\T \gamma^*} (X_i-X_l) \sum_{k=1}^n R_{jk} \me^{X_k^\T \gamma^*} (X_i-X_k)^\T,
\end{align*}
where the last equality holds because $  p_j(X_k)  \me^{X_i^\T \gamma^*}  = p_j(X_k)  \me^{X_k^\T \gamma^*}$ under model (\ref{eq:prob-model2}).
Comparing the preceding two displays yields the desired result.

(ii) It suffices to show that if $\sum_{l=1}^n R_{jl} D_{jl} =1$, then
\begin{align*}
 \hat v_j(\gamma) =
 \left(\sum_{l=1}^n R_{jl}D_{jl} \right) \sum_{i=1}^n \frac{R_{ji} \me^{X_i^\T\gamma}}{\sum_{l=1}^n R_{jl} \me^{ \tilde X_l^\T \gamma} }
 \left( X_i - \frac{\sum_{l=1}^n R_{jl} \me^{\tilde X_l^\T \gamma} X_l } {\sum_{l=1}^n
 R_{jl} \me^{ X_l^\T \gamma}}  \right)^{\otimes 2} ,
\end{align*}
that is,
\begin{align}
& \sum_{i=1}^n R_{ji} (1-D_{ji}) \me^{X_i^\T \gamma}  \sum_{l=1}^n R_{jl} \me^{X_l^\T \gamma} (X_i-X_l) \sum_{k=1}^n R_{jk} D_{jk} (X_i-X_k)^\T  \nonumber \\
& = \left(\sum_{l=1}^n R_{jl}D_{jl} \right) \sum_{i=1}^n \frac{R_{ji} \me^{X_i^\T\gamma}}{\sum_{l=1}^n R_{jl} \me^{ \tilde X_l^\T \gamma} } \left\{ \sum_{l=1}^n R_{jl}\me^{X_l^\T \gamma} (X_i-X_l) \right\}^{\otimes 2} . \label{eq:prf-gam-model-based}
\end{align}
First, (\ref{eq:prf-gam-model-based}) holds trivially in the case where $\sum_{l=1}^n R_{jl} D_{jl} =1$, i.e., $D_{ji}=0$ for all $i \in I_j$,
where $I_j = \{i: R_{ji}=1,\, i=1,\ldots,n\}$.
Next suppose that $\sum_{l=1}^n R_{jl} D_{jl} =1$, i.e., there exists only one element $i_0 \in I_j$ such that $D_{ji_0}=1$ and $D_{ji}=0$ for $i\not=i_0$ and $i\in I_j$.
The left hand side of (\ref{eq:prf-gam-model-based}) can be calculated as
\begin{align*}
& \sum_{i=1}^n R_{ji} \me^{X_i^\T \gamma}  \sum_{l=1}^n R_{jl} \me^{X_l^\T \gamma} (X_i-X_l) (X_i-X_{i_0})^\T \\
& =  \sum_{i=1}^n R_{ji} \me^{X_i^\T \gamma}  \sum_{l=1}^n R_{jl} \me^{X_l^\T \gamma} (X_i-X_l) X_i^\T -  \sum_{i=1}^n R_{ji} \me^{X_i^\T \gamma}  \sum_{l=1}^n R_{jl} \me^{X_l^\T \gamma} (X_i-X_l) X_{i_0}^\T\\
& = \sum_{i=1}^n R_{ji} \me^{X_i^\T \gamma}  \sum_{l=1}^n R_{jl} \me^{X_l^\T \gamma} (X_i-X_l) X_i^\T,
\end{align*}
where the last equality holds because $\sum_{i=1}^n R_{ji} \me^{X_i^\T \gamma}  \sum_{l=1}^n R_{jl} \me^{X_l^\T \gamma} (X_i-X_l) =0$.
The right hand side of (\ref{eq:prf-gam-model-based}) can be calculated as
\begin{align*}
& \sum_{i=1}^n \frac{R_{ji} \me^{X_i^\T\gamma}}{\sum_{l=1}^n R_{jl} \me^{ \tilde X_l^\T \gamma} } \left\{ \sum_{l=1}^n R_{jl}\me^{X_l^\T \gamma} (X_i-X_l) \right\}^{\otimes 2} \\
& =\sum_{i=1}^n \frac{R_{ji} \me^{X_i^\T\gamma}}{\sum_{l=1}^n R_{jl} \me^{ \tilde X_l^\T \gamma} } \left\{ \sum_{l=1}^n R_{jl}\me^{X_l^\T \gamma} (X_i-X_l) \right\}  \left\{ \sum_{l=1}^n R_{jl}\me^{X_l^\T \gamma} X_i^\T \right\} \\
& = \sum_{i=1}^n  R_{ji} \me^{X_i^\T \gamma} X_i X_i^\T \sum_{l=1}^n R_{jl}\me^{X_l^\T \gamma} - \sum_{l=1}^n R_{jl}\me^{X_l^\T \gamma} X_l \sum_{i=1}^n R_{ji}\me^{X_i^\T \gamma} X_i^\T ,
\end{align*}
where the second equality holds because $\sum_{i=1}^n R_{ji} \me^{X_i^\T\gamma} \sum_{l=1}^n R_{jl}\me^{X_l^\T \gamma} (X_i-X_l) =0$.
Comparing the last two displays establishes (\ref{eq:prf-gam-model-based}).
{\hfill $\Box$ \vspace{.1in}}

\noindent\textbf{Calculation of $H(\beta)$ in Section~\ref{sec:odds-model}}.\;
The matrix $H(\beta)$ is defined as $-\partial\tau_\bullet(\beta)/\partial \beta^\T$, where $\tau_\bullet(\beta)$ is \
the right hand side of (\ref{eq:bet-est-pop}) or (\ref{eq:cmh-est-pop}), that is,
\begin{align*}
\tau_\bullet(\beta) =  \sum_{j=1}^J \left[ \frac{ E(R_j D_j X) E \{ \tilde R_j (1-\tilde D_j) \me^{ \tilde X^\T \beta} \} }{E(\tilde R_j \me^{ \tilde X^\T \beta})}
- \frac{ E\{R_j (1-D_j)  \me^{X^\T \beta} X\} E( \tilde R_j \tilde D_j) }{E(\tilde R_j \me^{ \tilde X^\T \beta})} \right]  .
\end{align*}
Taking derivatives of the two terms inside $\sum_{j=1}^J$ separately, we find
\begin{align*}
& \frac{\partial}{\partial \beta^\T}  \frac{ E(R_j D_j X) E \{ \tilde R_j (1-\tilde D_j) \me^{ \tilde X^\T \beta} \} }{E(\tilde R_j \me^{ \tilde X^\T \beta})}  \\
& = \frac{ E(R_j D_j X) E \{ \tilde R_j (1-\tilde D_j) \me^{ \tilde X^\T \beta} ( \tilde X^\T - \frac{E( R_j \me^{  X^\T \beta}  X^\T) }{ E( R_j \me^{  X^\T \beta})} ) \} }
{E(\tilde R_j \me^{ \tilde X^\T \beta})},
\end{align*}
and
\begin{align*}
& \frac{\partial}{\partial \beta^\T}  \frac{ E\{R_j (1-D_j)  \me^{X^\T \beta} X\} E( \tilde R_j \tilde D_j) }{E(\tilde R_j \me^{ \tilde X^\T \beta})}\\
& =  \frac{  E( \tilde R_j \tilde D_j) E\{R_j (1-D_j)  \me^{X^\T \beta} X (  X^\T - \frac{E(\tilde R_j \me^{ \tilde X^\T \beta} \tilde X^\T) }{ E(\tilde R_j \me^{ \tilde X^\T \beta})} ) \} }
{E(\tilde R_j \me^{ \tilde X^\T \beta})}
\end{align*}
Combining the two displays yields the stated expression for $H(\beta)$.
{\hfill $\Box$ \vspace{.1in}}

\noindent\textbf{Calculation of $G(\beta^*)$ in Section~\ref{sec:odds-model}}.\;
We show that for $j=1,\ldots,J$,
\begin{align*}
& \var \{g_{j1}( Y, \delta, X; \beta^*) \} \\
& = E\left[ R_j \me^{\beta^*_{0j}+X^\T \beta^*} \frac{E^2(\tilde R_j(1-\tilde D_j) \me^{ \tilde X^\T \beta^*})} {E^2(\tilde R_j \me^{ \tilde X^\T \beta^*}) }
 \left\{ X - \frac{E( \tilde R_j (1-\tilde D_j)\me^{\tilde X^\T \beta^*} \tilde X )} {E(\tilde R_j (1-\tilde D_j)\me^{ \tilde X^\T \beta^*})}  \right\}^{\otimes 2} \right].
\end{align*}
Under model (\ref{eq:odds-model2}), $E\{ g_{j1}( Y, \delta, X; \beta^*) | R_j,X \}=0$. Then it suffices to show that
\begin{align}
& \var \left\{  D_j E(\tilde R_j (1-\tilde D_j)\me^{ \tilde X^\T \beta^*}) -  (1-D_j)\me^{X^\T \beta^*} E( \tilde R_j\tilde D_j) | R_j=1, X \right\} \nonumber \\
& = \me^{\beta^*_{0j}+X^\T \beta^*} E^2(\tilde R_j(1-\tilde D_j) \me^{ \tilde X^\T \beta^*}) . \label{eq:bet-G-calc}
\end{align}
By direct calculation, we find
\begin{align*}
& \var \left\{  D_j -  (1-D_j)\me^{\beta^*_{0j}+X^\T \beta^*}  | R_j=1, X \right\} \\
& = (1+\me^{\beta^*_{0j}+X^\T \beta^*})^2 \var \left\{ D_j - \expit(\beta^*_{0j}+X^\T \beta^*) \right\}  = \me^{\beta^*_{0j}+X^\T \beta^*} .
\end{align*}
Substituting $\me^{\beta^*_{0j}} = E( \tilde R_j\tilde D_j) /E(\tilde R_j (1-\tilde D_j)\me^{ \tilde X^\T \beta^*}) $ into the left hand side above and rearranging yields the desired result.
{\hfill $\Box$ \vspace{.1in}}

\noindent\textbf{Proof of Proposition~\ref{pro:bet-model-robust}}.\;
It suffices to show that for $j=1,\ldots,J$,
\begin{align*}
& \hat\tau_j(\bar\beta) = \frac{1}{n} \sum_{i=1}^n g_j (Y_i,\delta_i,X_i ; \bar\beta) + o_p(n^{-1/2}),
\end{align*}
that is,
\begin{align}
&   \frac{1}{n} \sum_{i=1}^n R_{ji} \frac{ D_{ji} \sum_{l=1}^n R_{jl} (1-D_{jl})\me^{ X_l^\T \bar\beta} - (1-D_{ji}) \me^{X_i^\T \bar\beta}\sum_{l=1}^n R_{jl} D_{jl}}
{ \sum_{l=1}^n R_{jl} \me^{ X_l^\T \bar\beta} }  X_i \nonumber \\
& = \frac{1}{n} \sum_{i=1}^n  R_{ji} \frac{D_{ji} E(\tilde R_j (1-\tilde D_j)\me^{ \tilde X^\T \bar\beta}) -  (1-D_{ji})\me^{X_i^\T \bar\beta} E( \tilde R_j\tilde D_j)} {E(\tilde R_j \me^{ \tilde X^\T \bar\beta})}
\left\{ X_i - \frac{ E( \tilde R_j (1-\tilde D_j) \me^{\tilde X^\T \bar\beta}\tilde X)} {E( \tilde R_j (1-\tilde D_j) \me^{\tilde X^\T \bar\beta})} \right\} \nonumber \\
& \quad - \frac{E(\tilde R_j \tilde D_j \tilde X) E(\tilde R_j (1-\tilde D_j)\me^{ \tilde X^\T \bar\beta}) -  E(\tilde R_j(1-\tilde D_j)\me^{\tilde X^\T \bar\beta}\tilde X) E( \tilde R_j\tilde D_j)} { E(\tilde R_j \me^{ \tilde X^\T \bar\beta}) } \nonumber \\
& \quad \times \left\{ \frac{ R_{ji}  \me^{ X_i^\T \bar\beta}}{E(\tilde R_j \me^{ \tilde X^\T \bar\beta})} - \frac{ R_{ji} (1-D_{ji}) \me^{X_i^\T \bar\beta} }{E( \tilde R_j (1-\tilde D_j) \me^{\tilde X^\T \bar\beta})} \right\} + o_p(n^{-1/2}). \label{eq:prf1-bet-model-robust}
\end{align}
We use the expansions
\begin{align*}
& \frac{ \sum_{l=1}^n R_{jl} (1-D_{jl})\me^{ X_l^\T \bar\beta} }{ \sum_{l=1}^n R_{jl} \me^{ X_l^\T \bar\beta} } - \frac{ E ( \tilde R_j (1-\tilde D_j)\me^{\tilde X^\T \bar\beta} )}{ E(\tilde R_j \me^{\tilde X^\T \bar\beta} ) } \\
& = \frac{n^{-1}
\sum_{l=1}^n R_{jl} (1-D_{jl})\me^{ X_l^\T \bar\beta} - R_{jl} \me^{ X_l^\T \bar\beta}  \frac{ E ( \tilde R_j (1-\tilde D_j)\me^{\tilde X^\T \bar\beta} )}{ E(\tilde R_j \me^{\tilde X^\T \bar\beta} ) } }
{ E(\tilde R_j \me^{\tilde X^\T \bar\beta} )  } + o_p(n^{-1/2}) ,
\end{align*}
and
\begin{align*}
& \frac{ \sum_{l=1}^n R_{jl} D_{jl} }{ \sum_{l=1}^n R_{jl} \me^{ X_l^\T \bar\beta} } - \frac{ E ( \tilde R_j \tilde D_j )}{ E(\tilde R_j \me^{\tilde X^\T \bar\beta} ) } \\
& = \frac{ n^{-1}
\sum_{l=1}^n R_{jl} D_{jl}  - R_{jl} \me^{ X_l^\T \bar\beta}  \frac{ E ( \tilde R_j \tilde D_j )}{ E(\tilde R_j \me^{\tilde X^\T \bar\beta} ) } }
{ E(\tilde R_j \me^{\tilde X^\T \bar\beta} ) } + o_p(n^{-1/2}) .
\end{align*}
The left hand side of (\ref{eq:prf1-bet-model-robust}) can be approximated as
\begin{align}
& \frac{1}{n} \sum_{i=1}^n  R_{ji} \frac{D_{ji} E(\tilde R_j (1-\tilde D_j)\me^{ \tilde X^\T \bar\beta}) -  (1-D_{ji})\me^{X_i^\T \bar\beta} E( \tilde R_j\tilde D_j)} {E(\tilde R_j \me^{ \tilde X^\T \bar\beta})} X_i
\nonumber  \\
& \quad +  E(\tilde R_j \tilde D_j \tilde X)
\frac{ n^{-1} \sum_{l=1}^n R_{jl} (1-D_{jl})\me^{ X_l^\T \bar\beta} - R_{jl} \me^{ X_l^\T \bar\beta}  \frac{ E ( \tilde R_j (1-\tilde D_j)\me^{\tilde X^\T \bar\beta} )}{ E(\tilde R_j \me^{\tilde X^\T \bar\beta} ) } }
{ E(\tilde R_j \me^{\tilde X^\T \bar\beta} )  } \nonumber \\
& \quad -  E(\tilde R_j (1-\tilde D_j) \me^{\tilde X^\T \bar\beta} \tilde X)
\frac{ n^{-1} \sum_{l=1}^n R_{jl} D_{jl}  - R_{jl} \me^{ X_l^\T \bar\beta}  \frac{ E ( \tilde R_j \tilde D_j )}{ E(\tilde R_j \me^{\tilde X^\T \bar\beta} ) } }
{ E(\tilde R_j \me^{\tilde X^\T \bar\beta} ) } + o_p(n^{-1/2})  . \label{eq:prf2-bet-model-robust}
\end{align}
The sum of the last two terms of (\ref{eq:prf2-bet-model-robust}) can be decomposed as
\begin{align}
& E(\tilde R_j \tilde D_j \tilde X)\frac{ n^{-1} \sum_{l=1}^n R_{jl} (1-D_{jl})\me^{ X_l^\T \bar\beta} }
{ E(\tilde R_j \me^{\tilde X^\T \bar\beta} )  } -  E(\tilde R_j (1-\tilde D_j) \me^{\tilde X^\T \bar\beta} \tilde X)\frac{ n^{-1} \sum_{l=1}^n R_{jl} D_{jl} }
{ E(\tilde R_j \me^{\tilde X^\T \bar\beta} ) } \nonumber \\
& \quad - E(\tilde R_j \tilde D_j \tilde X) \frac{1}{n} \sum_{l=1}^n  R_{jl} \me^{ X_l^\T \bar\beta}  \frac{ E ( \tilde R_j (1-\tilde D_j)\me^{\tilde X^\T \bar\beta} )}{ E^2(\tilde R_j \me^{\tilde X^\T \bar\beta} ) }
\nonumber \\
& \quad + E(\tilde R_j (1-\tilde D_j) \me^{\tilde X^\T \bar\beta} \tilde X)\frac{1}{n} \sum_{l=1}^n  R_{jl} \me^{ X_l^\T \bar\beta}  \frac{ E ( \tilde R_j \tilde D_j )}{ E^2(\tilde R_j \me^{\tilde X^\T \bar\beta} ) }.
\label{eq:prf3-bet-model-robust}
\end{align}
Hence as a byproduct, (\ref{eq:prf2-bet-model-robust}) can be rewritten as
\begin{align*}
& \frac{1}{n} \sum_{i=1}^n  R_{ji} \frac{D_{ji} E(\tilde R_j (1-\tilde D_j)\me^{ \tilde X^\T \bar\beta}) }  {E(\tilde R_j \me^{ \tilde X^\T \bar\beta})}
\left\{ X_i - \frac{ E( \tilde R_j (1-\tilde D_j) \me^{\tilde X^\T \bar\beta}\tilde X)} {E( \tilde R_j (1-\tilde D_j) \me^{\tilde X^\T \bar\beta})} \right\} \\
& \quad - \frac{1}{n} \sum_{i=1}^n  R_{ji} \frac{(1-D_{ji})\me^{X_i^\T \bar\beta} E( \tilde R_j\tilde D_j)} {E(\tilde R_j \me^{ \tilde X^\T \bar\beta})}
\left\{ X_i - \frac{ E ( \tilde R_j \tilde D_j \tilde X)}{ E(\tilde R_j \tilde D_j) }  \right\} \\
& \quad - \frac{E(\tilde R_j \tilde D_j \tilde X) E(\tilde R_j (1-\tilde D_j)\me^{ \tilde X^\T \bar\beta}) -  E(\tilde R_j(1-\tilde D_j)\me^{\tilde X^\T \bar\beta}\tilde X) E( \tilde R_j\tilde D_j)}
{ E^2(\tilde R_j \me^{ \tilde X^\T \bar\beta}) }  \frac{1}{n} \sum_{i=1}^n R_{ji}  \me^{ X_i^\T \bar\beta} .
\end{align*}
The first term of (\ref{eq:prf3-bet-model-robust}) can be rearranged as
\begin{align}
& \frac{1}{n} \sum_{i=1}^n  R_{ji} \frac{ (1-D_{ji})\me^{X_i^\T \bar\beta} E( \tilde R_j\tilde D_j)} {E(\tilde R_j \me^{ \tilde X^\T \bar\beta})}
\cdot \frac{ E( \tilde R_j (1-\tilde D_j) \me^{\tilde X^\T \bar\beta}\tilde X)} {E( \tilde R_j (1-\tilde D_j) \me^{\tilde X^\T \bar\beta})} \nonumber \\
& + \left\{ E(\tilde R_j \tilde D_j \tilde X) - \frac{ E( \tilde R_j (1-\tilde D_j) \me^{\tilde X^\T \bar\beta}\tilde X)} {E( \tilde R_j (1-\tilde D_j) \me^{\tilde X^\T \bar\beta})}
 E(\tilde R_j \tilde D_j ) \right\} \frac{ n^{-1} \sum_{l=1}^n R_{jl} (1-D_{jl})\me^{ X_l^\T \bar\beta} }
{ E(\tilde R_j \me^{\tilde X^\T \bar\beta} )  } .  \label{eq:prf4-bet-model-robust}
\end{align}
Combining the first term of (\ref{eq:prf2-bet-model-robust}) and all terms of (\ref{eq:prf3-bet-model-robust}) and substituting (\ref{eq:prf4-bet-model-robust})
for the first term of (\ref{eq:prf3-bet-model-robust})
yields the right hand side of (\ref{eq:prf1-bet-model-robust}) as desired.
{\hfill $\Box$ \vspace{.1in}}

\noindent\textbf{Proof of Proposition~\ref{pro:bet-model-based}}.

(i) Both (\ref{eq:bet-unbiased-mean}) and (\ref{eq:bet-unbiased-var}) follow from Proposition~\ref{pro:bet-model-based2}.

(ii) The result holds trivially in the case where $D_{ji}=0$ for all $i \in I_j$, where
$I_j = \{i: R_{ji}=1,\, i=1,\ldots,n\}$.
Suppose that there exists only one element $i_0 \in I_j$ such that $D_{ji_0}=1$ and $D_{ji}=0$ for $i\not=i_0$ and $i\in I_j$.

First, we show that $\hat H(\beta) = \hat B(\beta)$. It suffices to demonstrate
\begin{align}
&  \sum_{i=1}^n  \frac{R_{ji} \me^{X_i^\T \beta}} {\sum_{l=1}^n  R_{jl} \me^{ X_l^\T \beta} } (X_i-X_{i_0})
\left( X_i^\T - \frac{\sum_{l=1}^n  R_{jl} \me^{ X_l^\T \beta} X_l^\T }{\sum_{l=1}^n R_{jl} \me^{ X_l^\T \beta} } \right) \nonumber  \\
&= \sum_{i=1}^n \frac{R_{ji} \me^{X_i^\T\gamma}}{\sum_{l=1}^n R_{jl} \me^{ X_l^\T \gamma} } \left( X_i - \frac{\sum_{l=1}^n R_{jl} \me^{X_l^\T \gamma} X_l } {\sum_{l=1}^n
 R_{jl} \me^{ X_l^\T \gamma}}  \right)^{\otimes 2}  . \label{eq:prf1-bet-model-based-ii}
\end{align}
This follows because the value $X_{i_0}$ on the left hand side can be equivalently replaced by $\sum_{l=1}^n  R_{jl} \me^{ X_l^\T \beta} X_l^\T /\sum_{l=1}^n R_{jl} \me^{ X_l^\T \beta}$ using the fact that
\begin{align*}
&  \sum_{i=1}^n  \frac{R_{ji} \me^{X_i^\T \beta}} {\sum_{l=1}^n  R_{jl} \me^{ X_l^\T \beta} }
\left( X_i^\T - \frac{\sum_{l=1}^n  R_{jl} \me^{ X_l^\T \beta} X_l^\T }{\sum_{l=1}^n R_{jl} \me^{ X_l^\T \beta} } \right)   =0.
\end{align*}

Second, we show that $\hat G_{\text{b2}} (\beta) = \hat B(\beta)$. It suffices to demonstrate
\begin{align*}
&  \frac{\sum_{i=1}^n R_{ji} \me^{X_i^\T\beta + X_{i_0}^\T \beta} (X_i - X_{i_0})^{\otimes 2} +
\sum_{i=1}^n R_{ji} \me^{X_i^\T\beta} \sum_{l\not= i_0} R_{jl} \me^{X_l^\T \beta} (X_i - X_l) (X_i - X_{i_0})^\T }{(\sum_{l=1}^n R_{jl} \me^{ X_l^\T \beta})^2}   \\
&= \sum_{i=1}^n  \frac{R_{ji} \me^{X_i^\T\gamma}}{\sum_{l=1}^n R_{jl} \me^{ X_l^\T \gamma} } \left( X_i - \frac{\sum_{l=1}^n R_{jl} \me^{X_l^\T \gamma} X_l } {\sum_{l=1}^n
 R_{jl} \me^{ X_l^\T \gamma}}  \right)^{\otimes 2}  ,
\end{align*}
that is,
\begin{align}
&  \frac{\sum_{i=1}^n R_{ji} \me^{X_i^\T\beta} \sum_{l=1}^n R_{jl}\me^{X_l^\T \beta} (X_i - X_l) (X_i - X_{i_0})^\T }{(\sum_{l=1}^n R_{jl} \me^{ X_l^\T \beta})^2}  \nonumber  \\
&= \sum_{i=1}^n  \frac{R_{ji} \me^{X_i^\T\gamma}}{\sum_{l=1}^n R_{jl} \me^{ X_l^\T \gamma} } \left( X_i - \frac{\sum_{l=1}^n R_{jl} \me^{X_l^\T \gamma} X_l } {\sum_{l=1}^n
 R_{jl} \me^{ X_l^\T \gamma}}  \right)^{\otimes 2}  ,\label{eq:prf2-bet-model-based-ii}
\end{align}
But the left hand side of (\ref{eq:prf2-bet-model-based-ii}) is the transpose of that of (\ref{eq:prf1-bet-model-based-ii}) and hence
the desired result follows from (\ref{eq:prf1-bet-model-based-ii}).
{\hfill $\Box$ \vspace{.1in}}

\noindent\textbf{Proof of Proposition~\ref{pro:bet-model-based2}}.\;
We repeatedly use the fact that under model (\ref{eq:odds-model2}),
\begin{align}
& E \left\{ R_{ji} R_{jl} D_{ji} (1-D_{jl}) \me^{ X_l^\T \beta^*}  | T_j, R_{j,1:n}, X_{1:n} \right\} \nonumber \\
& = E \left\{ R_{ji} R_{jl} D_{jl} (1-D_{ji})  \me^{ X_i^\T \beta^*}  | T_j, R_{j,1:n}, X_{1:n} \right\}, \quad i,l=1,\ldots,n . \label{eq:prf1-bet-model-based2}
\end{align}
By equivalence between (\ref{eq:bet-est}) and (\ref{eq:cmh-est}), rewrite $\hat\tau_j(\beta^*)$ as
\begin{align*}
& \hat\tau_j(\beta^*) = (n \mu_j)^{-1}  \left(\sum_{i<l} +  \sum_{i>l} \right) R_{ji}R_{jl} D_{ji} (1-D_{jl})  \me^{ X_l^\T \beta^*}  (X_i-X_l) \\
& = (n \mu_j)^{-1}  \sum_{i <l} R_{ji}R_{jl} \left\{ D_{ji} (1-D_{jl})  \me^{ X_l^\T \beta^*} -  D_{jl} (1-D_{ji})  \me^{ X_i^\T \beta^*} \right\} (X_i - X_l),
\end{align*}
where $\mu_j  = \sum_{l=1}^n R_{jl} \me^{ X_l^\T \beta^*}$.
Then $E\{ \hat\tau_j(\beta^*) | T_j, R_{j,1:n}, X_{1:n}\} =0$ directly from (\ref{eq:prf1-bet-model-based2}).

Next, we calculate the conditional variance of $\hat\tau_j(\beta^*)$, that is,
$\var\{ \hat\tau_j(\beta^*) | T_j, R_{j,1:n}, X_{1:n}\}$ $ = E\{ \hat\tau_j^2 (\beta^*) | T_j, R_{j,1:n}, X_{1:n}\}$, where
\begin{align}
\hat\tau_j^2 (\beta^*)
& = (n \mu_j)^{-2}  \sum_{i <l}  \sum_{r<k} R_{ji}R_{jl} \left\{ D_{ji} (1-D_{jl})  \me^{ X_l^\T \beta^*} -  D_{jl} (1-D_{ji})  \me^{ X_i^\T \beta^*} \right\} (X_i - X_l) \times \nonumber  \\
& \quad R_{jr}R_{jk} \left\{ D_{jr} (1-D_{jk})  \me^{ X_k^\T \beta^*} -  D_{jk} (1-D_{jr})  \me^{ X_r^\T \beta^*} \right\} (X_r - X_k)^\T . \label{eq:prf2-bet-model-based2}
\end{align}
If neither of $(i,l)$ is included in $(r,k)$, then the conditional expectation of the corresponding term is 0 by (\ref{eq:prf1-bet-model-based2}).
The sum of the remaining terms in (\ref{eq:prf2-bet-model-based2}) can be decomposed as
$W_1 + W_2 + W_2^\T + W_3 + W_3^\T + W_4 + W_4^\T$, where
\begin{align*}
W_1 &= \sum_{i <l} R_{ji} R_{jl}  \left\{ D_{ji} (1-D_{jl})  \me^{ X_l^\T \beta^*} -  D_{jl} (1-D_{ji})  \me^{ X_i^\T \beta^*} \right\}^2 (X_i - X_l)^{\otimes2}\\
& = \sum_{i <l} R_{ji} R_{jl}  \left\{ D_{ji} (1-D_{jl})  \me^{ 2X_l^\T \beta^*} +  D_{jl} (1-D_{ji})  \me^{ 2X_i^\T \beta^*} \right\} (X_i - X_l)^{\otimes2} \\
& = \sum_{i ,l} R_{ji} R_{jl} D_{ji} (1-D_{jl})  \me^{ 2X_l^\T \beta^*}  (X_i - X_l)^{\otimes2},
\end{align*}
\begin{align*}
W_2 &= \sum_{i <l<k} R_{ji} R_{jl}R_{jk}  \left\{ D_{ji} (1-D_{jl})  \me^{ X_l^\T \beta^*} -  D_{jl} (1-D_{ji})  \me^{ X_i^\T \beta^*} \right\} (X_i - X_l) \times \\
& \quad \left\{ D_{ji} (1-D_{jk})  \me^{ X_k^\T \beta^*} -  D_{jk} (1-D_{ji})  \me^{ X_i^\T \beta^*} \right\} (X_i - X_k)^\T  \\
& = \sum_{i <l<k} R_{ji} R_{jl}R_{jk}  \left\{ D_{ji} (1-D_{jl}) (1-D_{jk}) \me^{ (X_l+X_k)^\T \beta^*} +  D_{jl} D_{jk}(1-D_{ji})  \me^{ 2X_i^\T \beta^*} \right\} \times \\
& \quad (X_i - X_l)  (X_i - X_k)^\T,
\end{align*}
\begin{align*}
W_3 &= \sum_{i <l<k} R_{ji} R_{jl}R_{jk}  \left\{ D_{ji} (1-D_{jl})  \me^{ X_l^\T \beta^*} -  D_{jl} (1-D_{ji})  \me^{ X_i^\T \beta^*} \right\} (X_i - X_l) \times \\
& \quad \left\{ D_{jl} (1-D_{jk})  \me^{ X_k^\T \beta^*} -  D_{jk} (1-D_{jl})  \me^{ X_l^\T \beta^*} \right\} (X_l - X_k)^\T  \\
& = -\sum_{i <l<k} R_{ji} R_{jl}R_{jk}  \left\{ D_{jl} (1-D_{ji}) (1-D_{jk}) \me^{ (X_i+X_k)^\T \beta^*} +  D_{ji} D_{jk}(1-D_{jl})  \me^{ 2X_l^\T \beta^*} \right\} \times \\
& \quad (X_i - X_l)  (X_l - X_k)^\T,
\end{align*}
and
\begin{align*}
W_4 &= \sum_{i <l<k} R_{ji} R_{jl}R_{jk}  \left\{ D_{ji} (1-D_{jk})  \me^{ X_k^\T \beta^*} -  D_{jk} (1-D_{ji})  \me^{ X_i^\T \beta^*} \right\} (X_i - X_k) \times \\
& \quad \left\{ D_{jl} (1-D_{jk})  \me^{ X_k^\T \beta^*} -  D_{jk} (1-D_{jl})  \me^{ X_l^\T \beta^*} \right\} (X_l - X_k)^\T  \\
& = \sum_{i <l<k} R_{ji} R_{jl}R_{jk}  \left\{ D_{jk} (1-D_{ji}) (1-D_{jl}) \me^{ (X_i+X_l)^\T \beta^*} +  D_{ji} D_{jl}(1-D_{jk})  \me^{ 2X_k^\T \beta^*} \right\} \times \\
& \quad (X_i - X_k)  (X_l - X_k)^\T .
\end{align*}
By relabeling the indices, we find
\begin{align}
& W_1+ W_2 + W_2^\T + W_3 + W_3^\T + W_4 + W_4^\T \nonumber \\
& = \sum_{i ,l} R_{ji} R_{jl} \, D_{ji} (1-D_{jl})  \me^{ 2X_l^\T \beta^*}  (X_i - X_l)^{\otimes2} \nonumber \\
& \quad + \sum_i \sum_{l\not= k} R_{ji} R_{jl}R_{jk} \, D_{ji} (1-D_{jl}) (1-D_{jk}) \me^{ (X_l+X_k)^\T \beta^*} (X_i - X_l)  (X_i - X_k)^\T \nonumber \\
& \quad + \sum_i \sum_{l\not= k} R_{ji} R_{jl}R_{jk} \, (1-D_{ji}) D_{jl} D_{jk}  \me^{ 2X_i^\T \beta^*} (X_i - X_l)  (X_i - X_k)^\T , \label{eq:prf3-bet-model-based2}
\end{align}
denoted as $W_1 + W_5 + W_6$.
Then $n^{-1} \mu_j^{-2}$ times the right hand side of (\ref{eq:prf3-bet-model-based2}) is conditionally unbiased for the conditional variance of $n^{1/2} \hat\tau_j(\beta^*)$.

Finally, $\hat\sigma_j(\beta^*)$ can be expressed as
\begin{align}
& \sum_{i ,l} R_{ji} R_{jl} \, (1-D_{ji}) D_{jl}  \me^{ (X_i+X_l)^\T \beta^*}  (X_i - X_l)^{\otimes2}  \nonumber  \\
& \quad + \sum_i \sum_{l, k} R_{ji} R_{jl}R_{jk} \,  (1-D_{jl}) D_{jk} \me^{ (X_i+X_l)^\T \beta^*} (X_i - X_l)  (X_i - X_k)^\T \nonumber  \\
& = \sum_{i ,l} R_{ji} R_{jl} \, (1-D_{ji}) D_{jl}  \me^{ (X_i+X_l)^\T \beta^*}  (X_i - X_l)^{\otimes2} \nonumber  \\
& \quad + \sum_i \sum_{l, k} R_{ji} R_{jl}R_{jk} \, (1- D_{ji}) (1-D_{jl}) D_{jk} \me^{ (X_i+X_l)^\T \beta^*} (X_i - X_l)  (X_i - X_k)^\T \nonumber  \\
& \quad + \sum_i \sum_{l, k} R_{ji} R_{jl}R_{jk} \, D_{ji} (1-D_{jl}) D_{jk}  \me^{ (X_i+X_l)^\T \beta^*} (X_i - X_l)  (X_i - X_k)^\T , \label{eq:prf4-bet-model-based2}
\end{align}
denoted as $W_7 + W_8 + W_9$.
By using (\ref{eq:prf1-bet-model-based2}), we see that
the conditional expectations of $W_1$, $W_5$, and $W_6$ are equal to those of $W_7$, $W_8$, and $W_9$ respectively.
Hence $\hat\sigma_j(\beta^*)$ is conditionally unbiased for the conditional variance of $n^{1/2} \hat\tau_j(\beta^*)$.
{\hfill $\Box$ \vspace{.1in}}

\noindent\textbf{Conditional unbiasedness of $\tilde\sigma_j(\beta^*)$}.\;
We continue from the Proof of Proposition~\ref{pro:bet-model-based2}. By definition, $\hat\sigma_j(\beta^*)$ can be expressed as
\begin{align*}
& \sum_i \sum_{l,k} R_{ji} R_{jl}R_{jk} \, (1-D_{ji}) D_{jl} D_{jk}  \me^{ 2X_i^\T \beta^*} (X_i - X_l)  (X_i - X_k)^\T \\
&  + \sum_i \sum_{l, k} R_{ji} R_{jl}R_{jk} \, (1- D_{ji}) (1-D_{jl}) D_{jk} \me^{ (X_i+X_l)^\T \beta^*} (X_i - X_l)  (X_i - X_k)^\T  .
\end{align*}
The first term is equal to $W_1 + W_6$ in (\ref{eq:prf3-bet-model-based2}),
and the second term is equal to $W_8$ in (\ref{eq:prf4-bet-model-based2}).
But the conditional expectation of $W_5$ is equal to that of $W_8$. Hence
$\tilde\sigma_j(\beta^*)$ is conditionally unbiased for the conditional variance of $n^{1/2} \hat\tau_j(\beta^*)$.
{\hfill $\Box$ \vspace{.1in}}

\noindent\textbf{Symmetric expression of $\tilde \sigma_j(\beta)$}.\;
We show that for $j=1,\ldots,J$, $\tilde \sigma_j(\beta)$ defined in (\ref{eq:sigma-robins}) can be equivalently expressed as
\begin{align}
& \tilde \sigma_j(\beta) = \frac{1}{n}\sum_{i=1}^n R_{ji}  (1-D_{ji})  \me^{X_i^\T \beta} \times \nonumber \\
& \quad \frac{ \me^{X_i^\T \beta} \{\sum_{l=1}^n R_{jl} D_{jl} (X_i -X_l)\}^{\otimes 2} + \frac{\sum_{l=1}^n R_{jl} D_{jl}}{\sum_{l=1}^n R_{jl} (1-D_{jl}) \me^{X_l^\T \beta} } \{\sum_{l=1}^n R_{jl} (1-D_{jl}) \me^{X_l^\T \beta} (X_i -X_l) \}^{\otimes 2} } {( \sum_{l=1}^n R_{jl} \me^{X_l^\T \beta} )^2} , \label{eq:sigma-sym}
\end{align}
which is a symmetric, nonnegative-definite matrix.
In fact,  $\tilde \sigma_j(\beta)$ in (\ref{eq:sigma-robins}) can be decomposed as
\begin{align*}
& \tilde \sigma_j(\beta) = \frac{1}{n} \sum_{i=1}^n R_{ji}  (1-D_{ji})  \me^{X_i^\T \beta}
\left\{ \frac{ \me^{X_i^\T \beta} \sum_{l=1}^n R_{jl} D_{jl} (X_i -X_l)  \sum_{k=1}^n R_{jk} D_{jk} (X_i -X_k)  } {( \sum_{l=1}^n R_{jl} \me^{X_l^\T \beta} )^2} \right. \\
& \qquad \left. + \frac{ \sum_{l=1}^n R_{jl} (1-D_{jl}) \me^{X_l^\T \beta} (X_i -X_l)
\sum_{k=1}^n R_{jk} D_{jk} (X_i -X_k)^\T  } {( \sum_{l=1}^n R_{jl} \me^{X_l^\T \beta} )^2} \right\}.
\end{align*}
To prove (\ref{eq:sigma-sym}), it suffices to show that
\begin{align*}
&  \sum_{i=1}^n R_{ji}  (1-D_{ji})  \me^{X_i^\T \beta}
  \sum_{l=1}^n R_{jl} (1-D_{jl}) \me^{X_l^\T \beta} (X_i -X_l) \sum_{k=1}^n R_{jk} D_{jk} (X_i -X_k)^\T \\
& = \frac{\sum_{l=1}^n R_{jl} D_{jl}}{\sum_{l=1}^n R_{jl} (1-D_{jl}) \me^{X_l^\T \beta} }  \sum_{i=1}^n R_{ji}  (1-D_{ji})  \me^{X_i^\T \beta}\left\{ \sum_{l=1}^n R_{jl} (1-D_{jl}) \me^{X_l^\T \beta} (X_i -X_l) \right\}^{\otimes 2} ,
\end{align*}
that is,
\begin{align}
&  \sum_{i=1}^n R_{ji}  (1-D_{ji})  \me^{X_i^\T \beta}
  \sum_{l=1}^n R_{jl} (1-D_{jl}) \me^{X_l^\T \beta} (X_i -X_l) \sum_{k=1}^n R_{jk} D_{jk} (X_i -X_k)^\T  \nonumber  \\
& = \left(\sum_{l=1}^n R_{jl} D_{jl} \right)  \sum_{i=1}^n R_{ji}  (1-D_{ji})  \me^{X_i^\T \beta} \times \nonumber \\
& \quad \left\{ \sum_{l=1}^n R_{jl} (1-D_{jl}) \me^{X_l^\T \beta} (X_i -X_l) \right\}
\left\{ X_i - \frac{\sum_{l=1}^n R_{jl} (1-D_{jl}) \me^{X_l^\T \beta} X_l}{\sum_{l=1}^n R_{jl} (1-D_{jl}) \me^{X_l^\T \beta}}  \right\}. \label{eq:prf-sigma-sym}
\end{align}
The left hand side of (\ref{eq:prf-sigma-sym}) can be calculated as
\begin{align*}
& \sum_{i=1}^n R_{ji}  (1-D_{ji})  \me^{X_i^\T \beta} \times \\
& \quad \left\{ \sum_{l=1}^n R_{jl} (1-D_{jl}) \me^{X_l^\T \beta} X_i  \sum_{k=1}^n R_{jk} D_{jk} X_i^\T
 -  \sum_{l=1}^n R_{jl} (1-D_{jl}) \me^{X_l^\T \beta} X_i \sum_{k=1}^n R_{jk} D_{jk} X_k^\T  \right. \\
& \qquad \left. - \sum_{l=1}^n R_{jl} (1-D_{jl}) \me^{X_l^\T \beta} X_l  \sum_{k=1}^n R_{jk} D_{jk} X_i^\T
 +  \sum_{k=1}^n R_{jk} (1-D_{jk}) \me^{X_l^\T \beta} X_l \sum_{k=1}^n R_{jk}  D_{jk} X_k^\T \right\}.
\end{align*}
The two terms corresponding to the second and fourth terms in the curly brackets are canceled with each other. Hence the left hand side of (\ref{eq:prf-sigma-sym}) reduces to
\begin{align*}
&  \sum_{i=1}^n R_{ji}  (1-D_{ji})  \me^{X_i^\T \beta}
  \sum_{l=1}^n R_{jl} (1-D_{jl}) \me^{X_l^\T \beta} (X_i -X_l) \sum_{k=1}^n R_{jk} D_{jk} X_i^\T  \\
& =  \sum_{k=1}^n R_{jk} D_{jk}  \sum_{i=1}^n R_{ji}  (1-D_{ji})  \me^{X_i^\T \beta}\sum_{l=1}^n R_{jl} (1-D_{jl}) \me^{X_l^\T \beta} (X_i -X_l) X_i^\T ,
\end{align*}
which is identical to the right hand side of  (\ref{eq:prf-sigma-sym})  because
\begin{align*}
&  \sum_{k=1}^n R_{jk} D_{jk}  \sum_{i=1}^n R_{ji}  (1-D_{ji})  \me^{X_i^\T \beta}\sum_{l=1}^n R_{jl} (1-D_{jl}) \me^{X_l^\T \beta} (X_i -X_l)=0,
\end{align*}
by direct calculation.

\section{Additional simulation results}

Tables~\ref{tab:sim-contT-phF}--\ref{tab:sim-contF-phF} present additional results from 2000 repeated simulations, where
proportional hazards are violated.
For each simulation, a sample of size $n=200$ is generated similarly as in Section~\ref{sec:simulation},
except for the following changes.
The event time $\tilde T$ is generated as Weibull with shape parameter 4 scale parameter $\exp(- X^\T \beta^*)$,
or Weibull with shape parameter 1 and scale parameter $\exp(- X^\T \beta^*)$ in the standard treatment group.
The censoring time $\tilde C$ is generated as $4 \exp (- X^\T \beta^*)$ times Beta $(2,2)$ in the test treatment group,
or Uniform between 0 and $4 \exp (- X^\T \beta^*)$ in the standard treatment group.
Both probability model (\ref{eq:prob-model}) and odds model (\ref{eq:odds-model})
are fit with the regression terms \texttt{Tr} and \texttt{X1}--\texttt{X4}.
Similar conclusions can be drawn as in Section~\ref{sec:simulation}, although all the variance estimates
exhibit more substantial under-estimation.

\clearpage


\begin{table}
\caption{Comparison from simulated data (finely discretized, non-prop hazards)} \label{tab:sim-contT-phF}
\footnotesize
\begin{center} \renewcommand\arraystretch{.9}
\begin{tabular*}{.9\textwidth}{ll rrrrr rrrrr} \hline
    && BP  & Efron  &   CML  &  wMH    & Plogit   & BP  & Efron  &   CML  &  wMH    & Plogit   \\
    &&   \multicolumn{5}{c}{Point mean}  &   \multicolumn{5}{c}{Point SD} \\ \cline{2-12}

Tr && $-.563$ & $-.564$ & $-.570$ & $-.563$ & $-.613$ & $.303$ & $.304$ & $.307$ & $.306$ & $.323$ \\
X1 && $1.093$ & $1.099$ & $1.106$ & $1.106$ & $1.157$ & $.212$ & $.213$ & $.215$ & $.215$ & $.228$ \\
X2 && $-.732$ & $-.737$ & $-.741$ & $-.741$ & $-.775$ & $.205$ & $.207$ & $.208$ & $.208$ & $.218$ \\
X3 && $.546$ & $.549$ & $.553$ & $.553$ & $.578$ & $.186$ & $.188$ & $.189$ & $.189$ & $.198$ \\
X4 && $.184$ & $.186$ & $.187$ & $.187$ & $.195$ & $.154$ & $.155$ & $.156$ & $.156$ & $.162$ \\ [2ex] \hline

    && oldBP   & BP  & Efron  &   CML  &  wMH    & Plogit   & BP  & Efron  &  wMH    & Plogit   \\
    && \multicolumn{6}{c}{Model-based SE}  & \multicolumn{4}{c}{Model-robust SE}  \\       \cline{2-12}

Tr && $.244$ & $.243$ & $.244$ & $.246$ & $.245$ & $.253$ & $.250$ & $.251$ & $.253$ & $.272$ \\
X1 && $.179$ & $.178$ & $.179$ & $.181$ & $.181$ & $.187$ & $.188$ & $.189$ & $.191$ & $.204$ \\
X2 && $.175$ & $.174$ & $.175$ & $.176$ & $.177$ & $.182$ & $.174$ & $.175$ & $.176$ & $.187$ \\
X3 && $.168$ & $.166$ & $.168$ & $.169$ & $.169$ & $.174$ & $.163$ & $.164$ & $.165$ & $.175$ \\
X4 && $.143$ & $.142$ & $.143$ & $.144$ & $.144$ & $.148$ & $.135$ & $.136$ & $.137$ & $.145$ \\ \hline
\end{tabular*} 
\end{center} \vspace{-.2in}
\end{table}

\begin{table}
\caption{Comparison from simulated data (coarsely discretized, non-prop hazards)} \label{tab:sim-contF-phF}
\footnotesize
\begin{center} \renewcommand\arraystretch{.9}
\begin{tabular*}{.9\textwidth}{ll rrrrr rrrrr} \hline
    && BP  & Efron  &   CML  &  wMH    & Plogit   & BP  & Efron  &   CML  &  wMH    & Plogit   \\
    &&   \multicolumn{5}{c}{Point mean}  &   \multicolumn{5}{c}{Point SD} \\ \cline{2-12}

Tr && $-.505$ & $-.518$ & $-.654$ & $-.501$ & $-.692$ & $.265$ & $.296$ & $.352$ & $.335$ & $.366$ \\
X1 && $.949$ & $1.049$ & $1.189$ & $1.201$ & $1.234$ & $.180$ & $.205$ & $.243$ & $.268$ & $.254$ \\
X2 && $-.635$ & $-.703$ & $-.797$ & $-.806$ & $-.827$ & $.178$ & $.200$ & $.229$ & $.247$ & $.238$ \\
X3 && $.474$ & $.525$ & $.595$ & $.600$ & $.617$ & $.162$ & $.183$ & $.208$ & $.222$ & $.216$ \\
X4 && $.160$ & $.177$ & $.200$ & $.204$ & $.208$ & $.136$ & $.152$ & $.170$ & $.179$ & $.176$ \\ [2ex] \hline

    && oldBP   & BP  & Efron  &   CML  &  wMH    & Plogit   & BP  & Efron  &  wMH    & Plogit   \\
    && \multicolumn{6}{c}{Model-based SE}  & \multicolumn{4}{c}{Model-robust SE}  \\       \cline{2-12}

Tr && $.243$ & $.220$ & $.243$ & $.276$ & $.270$ & $.283$ & $.225$ & $.247$ & $.277$ & $.315$ \\
X1 && $.172$ & $.150$ & $.176$ & $.206$ & $.233$ & $.211$ & $.164$ & $.183$ & $.229$ & $.234$ \\
X2 && $.170$ & $.150$ & $.173$ & $.198$ & $.216$ & $.202$ & $.154$ & $.171$ & $.206$ & $.212$ \\
X3 && $.164$ & $.145$ & $.166$ & $.188$ & $.204$ & $.193$ & $.146$ & $.161$ & $.190$ & $.197$ \\
X4 && $.141$ & $.125$ & $.142$ & $.159$ & $.170$ & $.163$ & $.122$ & $.134$ & $.155$ & $.161$ \\ \hline
\end{tabular*} 
\end{center} \vspace{-.2in}
\end{table}

\end{document}